\documentclass{article}

\usepackage{PRIMEarxiv}

\usepackage[utf8]{inputenc} 
\usepackage[T1]{fontenc}    
\usepackage{hyperref}       
\usepackage{url}            
\usepackage{booktabs}       
\usepackage{amsfonts}       
\usepackage{nicefrac}       
\usepackage{microtype}      
\usepackage{lipsum}
\usepackage{fancyhdr}       
\usepackage{graphicx}       
\graphicspath{{media/}}     

\newcommand\Rey{\mathit{Re}}

\usepackage{float}
\usepackage{subfig}

\usepackage{color}
\usepackage[x11names]{xcolor}
\hypersetup{
	colorlinks = true,
	linkcolor=blue,
	citecolor=SpringGreen4,
}

\usepackage[numbers, sort, comma, square]{natbib}

\usepackage{amsmath,amssymb,amsfonts,latexsym}

\title{
Liouville models of particle-laden flow
}

\author{
  Daniel Dom\'inguez-V\'azquez\thanks{ddominguezvazquez@sdsu.edu}, \ 
  Gustaaf B. Jacobs\thanks{Author to whom correspondence should be addressed: gjacobs@sdsu.edu} \\
  Department of Aerospace Engineering \\
  San Diego State University \\
  San Diego, CA 92182, USA\\
   \And
  Daniel M. Tartakovsky\thanks{tartakovsky@stanford.edu} \\
  Department of Energy Science and Engineering \\
  Stanford University \\
  Stanford, CA 94305, USA\\
}

\begin{document}

\maketitle

\begin{abstract}
Langevin (stochastic differential) equations are routinely used to describe particle-laden flows. They predict Gaussian probability density functions (PDFs) of a particle's trajectory and velocity, even though experimentally observed dynamics might be highly non-Gaussian. Our Liouville approach overcomes this dichotomy by replacing the Wiener process in the Langevin models with a (small) set of random variables, whose distributions are tuned to match the observed statistics. 
This strategy gives rise to an exact 
(deterministic, first-order, hyperbolic) Liouville equation that describes the evolution of a joint PDF in the augmented phase-space spanned by the random variables and the particle position and velocity. 
Analytical PDF solutions for canonical models of particle-laden flows serve to establish the relationship between the Langevin and Liouville approaches. 
Finally, our framework is used to derive a new analytical PDF model for fluidized homogeneous heating systems.
\end{abstract}


\section{Introduction} 
\label{sec: Introduction}

Dispersion of particles in chaotic flows is of fundamental importance to a plethora of natural environments and engineering technologies. For example, mixing of evaporating fuel droplets in a combustor determines the fuel-to-air ratio, homogeneity of the mixture, and ultimately the efficiency of the combustor; upwelling of surface dust particles by shocked, unsteady nozzle-jet flows can damage launch vehicles; etc. An accurate model prediction of such flows at large scales is necessary to elucidate fundamental physics and improve engineering design.

A typical model of turbulent particle-laden flows involves the spatial scales ranging from tens of meters (the application scale) to microns (the smallest scales of turbulence and particulates). Consequently, a predictive simulation based on ``first principles'' is not practical and a portion of the smaller scales has to be treated empirically through stochastic ``subgrid-scale'' models, as it is done in, e.g., large-eddy simulations (LES) or Reynolds-averaged Navier-Stokes (RANS) models. In the context of particle-laden flows, this stochasticity stems from either turbulence of a carrier gas/liquid (subgrid velocity fluctuations) or from random particle kinetics. 

Classical kinetic theory \cite{buyevich1971statistical1, buyevich1972statistical2, buyevich1972statistical3} deals with stochasticity on the particle phase through an equation for the joint probability density function (PDF) in the six--dimensional phase space composed by particle position and velocity.
This equation requires a closure approximation, which can be derived either analytically or numerically \cite{reeks1980eulerian, reeks1983transport, reeks1991kinetic, reeks1992continuum, swailes1997generalized, garzo1999dense, garzo2012enskog, bragg2012drift, swailes1998chapman, zaichik2007refinement, alipchenkov2005dispersion, pandya2001probability, pandya2002turbulent, pandya2003non, reeks2021development}.
The high--dimensionality of this formulation leads to the so--called \text{curse of dimensionality}, which has inspired reduced formulations based on a number of moments of the joint PDF \cite{marchisio2002comparison, marchisio2002quadratic, marchisio2003quadraturea, marchisio2003quadratureb, heylmun2019quadrature, bryngelson2020gaussian, charalampopoulos2022hybrid, bryngelson2023conditional}.
 
Equivalent stochastic differential equations (SDEs), which in this field are referred to as a generalized Langevin model~\cite{reeks2021development}, can be used to reduce the dimensionality of this PDF equation. The model relies on Wiener increments and random walks to account for stochasticity in particle trajectories~\cite{pope1985pdf,haworth1986generalized}; 
more recent versions of such stochastic models of the dispersed particle phase 
have been proposed in the literature~\cite{iliopoulos2003stochastic,gao2004stochastic_a,gao2004stochastic_b,gao2004stochastic_c,shotorban2005modeling,shotorban2006stochastic_a,shotorban2006stochastic_b,sengupta2009spectral,pozorski2009filtered,pai2012two,garzo2012enskog,tenneti2016stochastic,esteghamatian2018stochastic,lattanzi2020stochastic,knorps2021stochastic,lattanzi2022stochastic,friedrich2022single,pietrzyk2022analysis}. A representative formulation of this Langevin approach, in $d$ spatial dimensions, describes the temporal evolution of an inertial point particle's location, $\boldsymbol{X}_\text{p}(t) \in \mathbb{R}^d$, and velocity, $\boldsymbol{U}_\text{p}(t) \in \mathbb{R}^d$, by adding Wiener increments to the particle position and velocity to Newton's second law: 
\begin{subequations}\label{eq:intro_Langevin}
\begin{align}
    \text{d}{\boldsymbol{X}_\text{p}} &= {\boldsymbol{U}_\text{p}} \text{d}t + \boldsymbol{b}_x(\boldsymbol{X}_\text{p},\boldsymbol{U}_\text{p},t) \, \text{d}\boldsymbol{W}_x ,  
    \label{eq:intro_Langevin_dX} \\ 
    \text{d}{\boldsymbol{U}_\text{p}} &= \boldsymbol{\mu}_{a}(\boldsymbol{X}_\text{p},\boldsymbol{U}_\text{p},t) \text{d}t + \boldsymbol{b}_u(\boldsymbol{X}_\text{p},\boldsymbol{U}_\text{p},t) \, \text{d}\boldsymbol{W}_u. 
    \label{eq:intro_Langevin_dU}
\end{align}
\end{subequations}
The particle's slow-varying average acceleration, $\boldsymbol{\mu}_a\in \mathbb{R}^d$, might depend in general on $\boldsymbol{U}_\text{p}$ and $\boldsymbol{X}_\text{p}$ through the evaluation of the bulk-flow velocity at the particle position, $\boldsymbol{u}(\boldsymbol{X}_\text{p},t)\in \mathbb{R}^d$ and the relative velocity of the two phases, time, and other model parameters. The particle's fast-varying dynamics is represented by the (random) Wiener process $\boldsymbol{W}_u(t)\in \mathbb{R}^m$, whose strength is encapsulated in the (deterministic) diffusion tensor $\boldsymbol{b}_u \in \mathbb{R}^{d\times m} $ that, in general, varies with $\boldsymbol{X}_\text{p}$, $\boldsymbol{U}_\text{p}$, and time $t$.
Similarly, Wiener increments $\boldsymbol{W}_x(t)\in \mathbb{R}^m$ are usually added to the particle position with diffusion tensor $\boldsymbol{b}_x \in \mathbb{R}^{d\times m}$ (see for example Refs.\cite{lattanzi2020stochastic} and \cite{sheikhi2007velocity}). 
The joint PDF of the particle position and velocity, $f_{\boldsymbol{X} \boldsymbol{U}}(\boldsymbol{x}_\text{p},\boldsymbol{u}_\text{p};t)$, in the case of Gaussian white-noise in~\eqref{eq:intro_Langevin}, 
satisfies the $(2d)$-dimensional Fokker-Planck equation
\begin{align}
\begin{split}
    \frac{\partial f_{\boldsymbol{X} \boldsymbol{U}}}{\partial t} &+ \nabla_{\boldsymbol{x}_\text{p}} \cdot \left( \boldsymbol{u}_\text{p} f_{\boldsymbol{X} \boldsymbol{U}} \right) 
    + \nabla_{\boldsymbol{u}_\text{p}} \cdot \left( \boldsymbol{\mu}_{a}(\boldsymbol{x}_\text{p},\boldsymbol{u}_\text{p},t) f_{\boldsymbol{X} \boldsymbol{U}} \right) = \nabla \cdot ( \boldsymbol{\mathcal D} \nabla f_{\boldsymbol{X} \boldsymbol{U}} ),
\end{split}    
\label{eq: intro_FokkerPlanck}
\end{align}
with the diffusion tensor $\boldsymbol{\mathcal D} = \boldsymbol{b}\boldsymbol{b}^\top/2 \in \mathbb{R}^{2d \times 2d}$ and $\nabla = (\nabla_{\boldsymbol{x}_\text{p}},\nabla_{\boldsymbol{u}_\text{p}} )^\top \in \mathbb{R}^{2d}$. 


At least two major challenges hinder the development of consistent Langevin models for particle-laden flows. First, since the Wiener process is Gaussian and white noise, it gives rise to the Gaussian PDF $f_{\boldsymbol{X} \boldsymbol{U}}(\boldsymbol{x}_\text{p},\boldsymbol{u}_\text{p};t)$. Yet, empirically observed (and numerically simulated) single-point statistics of particles are rarely Gaussian~\cite{gheorghiu2003power,biferale2012convection,capecelatro2015fluid,forgues2019gaussian,dominguez2021lagrangian},  which suggests that models like~\eqref{eq:intro_Langevin} provide \textit{ad hoc} (and possibly inaccurate) representations of subgrid particle dynamics. Second, closure models for diffusion coefficients in turbulent flows, let alone particle-laden turbulent flows, remain one of the open problems in fluid mechanics. In addition to having these foundational issues, the Langevin approach is also computationally expensive when Monte Carlo methods are used to estimate the tails of joint PDFs. 

A way to overcome these limitations is to replace the Wiener increments in~\eqref{eq:intro_Langevin} with random forcings, which are represented via series expansions involving proper sets of random variables~\cite{venturi2013exact,cho2014statistical,dominguez2021lagrangian,dominguez2023sparser,dominguez2023MoC}. 
This approach replaces (stochastic) Langevin equations~\eqref{eq:intro_Langevin} with a coupled system of $2d$ ordinary differential equations with random coefficients, 
\begin{subequations}\label{eq:intro_ODEs_random}
\begin{align}
    \frac{\text{d}{\boldsymbol{X}_\text{p}}}{\text{d}t} &= {\boldsymbol{U}}_\text{p} + \sum_{i=1}^N \boldsymbol{Z}_i \circ \boldsymbol{\varphi}_i(\boldsymbol{X}_\text{p},\boldsymbol{U}_\text{p},t),  \\ 
    \frac{\text{d}{\boldsymbol{U}_\text{p}}}{\text{d}t} &= \boldsymbol{\mu}_{a}(\boldsymbol{X}_\text{p},\boldsymbol{U}_\text{p},t)  + \sum_{i=1}^N \boldsymbol{\Xi}_i \circ \boldsymbol{\phi}_i(\boldsymbol{X}_\text{p},\boldsymbol{U}_\text{p},t).
\end{align}
\end{subequations}
Here, $\circ$ is the Hadamard product, $\boldsymbol{Z}_i \in\mathbb{R}^d$ and $\boldsymbol{\Xi}_i \in \mathbb{R}^d$ are the vectors of random coefficients in the $N$th-order polynomials with orthogonal bases $\boldsymbol{\varphi}_i \in \mathbb{R}^d $ and $\boldsymbol{\phi}_i \in \mathbb{R}^d$ respectively, defined deterministically. While the Wiener process, $\boldsymbol W_x(t)$ or $\boldsymbol W_u(t)$ in~\eqref{eq:intro_Langevin}, is a random \emph{function}, an infinite-dimensional object that necessitates the use of stochastic calculus, the parameters $\boldsymbol{Z}_i$ and $\boldsymbol{\Xi}_i$ are random \emph{variables}; hence, for any realization $\boldsymbol{z}_i$ and $\boldsymbol{\xi}_i$ of $\boldsymbol{Z}_i$ and $\boldsymbol{\Xi}_i$,~\eqref{eq:intro_ODEs_random} are deterministic differential equations amenable to a standard numerical or analytical treatment. Moreover, the joint PDF of the random inputs and outputs in~\eqref{eq:intro_ODEs_random} defined in an \textit{augmented phase space} that includes the random coefficients, $f_{\boldsymbol{X}\boldsymbol{U}\boldsymbol{\Xi}\boldsymbol{Z}}(\boldsymbol x_\text{p}, \boldsymbol u_\text{p}, \boldsymbol \xi, \boldsymbol z; t)$, satisfies the Liouville equation
\begin{align}
\begin{split}
    &\frac{\partial f_{\boldsymbol{X}\boldsymbol{U}\boldsymbol{\Xi}\boldsymbol{Z}}}{\partial t} + \nabla_{\boldsymbol{x}_\text{p}} \cdot [ ( \boldsymbol{u}_\text{p} + \sum_{i=1}^N \boldsymbol{z}_i \circ \boldsymbol{\varphi}_i(\boldsymbol{x}_\text{p},\boldsymbol{u}_\text{p},t) ) f_{\boldsymbol{X}\boldsymbol{U}\boldsymbol{\Xi}\boldsymbol{Z}} ] \\
    &+ \nabla_{\boldsymbol{u}_\text{p}} \cdot [ ( \boldsymbol{\mu}_a(\boldsymbol{x}_\text{p},\boldsymbol{u}_\text{p},t) + \sum_{i=1}^N \boldsymbol{\xi}_i \circ \boldsymbol{\phi}_i(\boldsymbol{x}_\text{p},\boldsymbol{u}_\text{p},t) ) f_{\boldsymbol{X}\boldsymbol{U}\boldsymbol{\Xi}\boldsymbol{Z}} ] = 0,
\end{split}
\label{eq:intro_Liouville}
\end{align}
a hyperbolic partial-differential equation that admits a solution via the method of characteristics (MoC) \cite{dominguez2023MoC}. This solution might or might not be Gaussian, depending on, e.g., the PDF of the initial data or the random coefficients.
Alternatively,~\eqref{eq:intro_ODEs_random}--\eqref{eq:intro_Liouville} can be used to learn the series in~\eqref{eq:intro_ODEs_random} from data, making the Liouville approach adequate for developing stochastic models using data assimilation techniques~\cite{dominguez2022adjoint}.
We point out that the functions $\boldsymbol{\varphi}_i$ and $\boldsymbol{\phi}_i$ do not need to act as basis functions necessarily for the formulation to be valid.

Our analysis pursues two goals. The first is to elucidate that the Liouville model of particle-laden flows, i.e., Eq.~\eqref{eq:intro_Liouville}, amounts to a generalization of the Langevin approach. 
We show it 
in two canonical examples in which the former reduces to the latter when the random variables $\boldsymbol{Z}_i$ and $\boldsymbol{\Xi}_i$ are Gaussian and the source terms are found to fit the solution of the Fokker-Planck equivalent model. An analytical generalization for non-Gaussian systems is provided with the Liouville model.
The second goal is to show the potential of the Liouville approach for developing stochastic models of the particle phase.
By adjusting the random coefficients to fit data, the Liouville approach can be framed in a data-driven manner. 
We propose an analytical model for fluidized homogeneous heating systems (FHHS) based on the Liouville approach. 
The accuracy of this model is established via comparison with the predictions of both particle-resolved direct numerical simulations (PR-DNS) and previous models proposed in the literature~\cite{koch1999particle,tenneti2016stochastic,lattanzi2020stochastic,lattanzi2022fluid,lattanzi2022stochastic}.


Section~\ref{sec: PDF_models} contains a brief formulation of both Langevin and Liouville frameworks for modeling the particle dynamics in multiphase flows. 
In Sections~\ref{sec: PDF_solutions}, we derive analytical solutions of the Fokker-Planck equation and the Liouville equation for canonical-flow examples and establish the equivalency between the Langevin and Liouville approaches. In Section~\ref{sec: FHHS}, our Liouville approach is used to derive a new model of homogeneous fluidized heated systems. Main conclusions drawn from this study are collated in Section~\ref{sec: conclusions}.


\section{Stochastic Models of Particle Dynamics} 
\label{sec: PDF_models}

If (the sum of) forces $\boldsymbol g(\boldsymbol x_\text{p}, \boldsymbol u_\text{p}, t)$ acting on a particle were known with certainty, then its trajectory $\boldsymbol x_\text{p}(t)$ and velocity $\boldsymbol u_\text{p}(t)$ would be deterministic and computable from Newton's second law, 
\begin{align}\label{eq: tracer_deterministic_dxdt}
\frac{ \text d \boldsymbol x_\text{p} }{  \text dt } = \boldsymbol u_\text{p}, \qquad
\frac{ \text d \boldsymbol u_\text{p} }{ \text dt } = \boldsymbol g.
\end{align}
In particle-laden flows, these forces are unresolved/uncertain, e.g., due to multi-particle interactions or drag on particles of non-canonical shapes. When treated probabilistically, this uncertainty gives rise to stochastic dynamics, such that the deterministic functions $\boldsymbol x_\text{p}(t)$ and velocity $\boldsymbol u_\text{p}(t)$ are replaced with the random processes $\boldsymbol X_\text{p}(t)$ and velocity $\boldsymbol U_\text{p}(t)$.

\subsection{Particle trajectory}
\label{sec: PDF_models_position}

A reduced-complexity model of particle dynamics superimposes stochastic fluctuations onto a given deterministic drift velocity $\boldsymbol u_\text{p}(t)$.
This model is used for example in the literature to compute Lagrangian Coherent Structures (LCS) of fluid particles for uncertain fluid velocity field~\cite{schneider2011variance}, also denoted particle Langevin model~\cite{lattanzi2020stochastic} for inertial particles with unresolved drag.


\subsubsection{Langevin approach}
\label{sec: PDF_models_position_Langevin}

For deterministic drift velocity $\boldsymbol{u}_\text{p}(t)$ and dispersion tensor $\boldsymbol{b}_x$, the Langevin approach is to replace~\eqref{eq: tracer_deterministic_dxdt} with stochastic ODEs~\cite{schneider2011variance,lattanzi2020stochastic} \begin{align}
    \text{d} {\boldsymbol{X}_\text{p}} = \boldsymbol{u}_\text{p} \text{d} t + \boldsymbol{b}_x({\boldsymbol{X}_\text{p}},t)\text{d}\boldsymbol{W}_x. 
    \label{eq: dXdt_langevin}
\end{align}
A single-point PDF of the stochastic trajectory $\boldsymbol X_\text{p}(t)$ in the Langevin equation~\eqref{eq: dXdt_langevin}, $f_{\boldsymbol{X}}(\boldsymbol{x}_\text{p};t)$, satisfies the Fokker-Planck equation \cite{risken1996fokker},
\begin{align}
    \frac{\partial f_{\boldsymbol{X}} }{\partial t} + \nabla_{\boldsymbol{x}_\text{p}} \cdot  \left( \boldsymbol{u}_\text{p} f_{\boldsymbol{X}}  \right) = \nabla_{\boldsymbol{x}_\text{p}} \cdot ( \boldsymbol{\mathcal D} \nabla_{\boldsymbol{x}_\text{p}} f_{\boldsymbol{X}} ), \quad
    \boldsymbol{\mathcal D} = \frac{ \boldsymbol{b}_x^{}  \, \boldsymbol{b}_x^\top }{2}.
    \label{eq: position_fokker_general}
\end{align}

\subsubsection{Liouville approach}
\label{sec: PDF_models_position_Liouville}

Instead of using the Wiener process $\boldsymbol W_x(t)$, the Liouville approach accounts for stochasticity of the particle dynamics using time-invariant random variables $\boldsymbol{Z}_i$, characterized by the joint PDF $f_{\boldsymbol{Z}}(\boldsymbol{z})$. Specifically, the stochastic ODE~\eqref{eq: dXdt_langevin} is replaced with an ODE with random coefficients,
\begin{align}
    \frac{\text{d} \boldsymbol{X}_\text{p}}{\text{d}t} = \boldsymbol{u}_\text{p} + \sum_{i=1}^{N}\boldsymbol{Z}_i \circ  \boldsymbol{\varphi}_i(\boldsymbol{X}_\text{p},t), 
    \label{eq: position_ODE_liouville_general}
\end{align}
where $\boldsymbol{\varphi}_i(\boldsymbol{X}_\text{p},t)$ is a prescribed vector function conforming a basis, for example Chebyshev polynomials. 
While the derivation of the PDF equation for $f_{\boldsymbol X}(\boldsymbol x_\text{p};t)$ requires a closure approximation~\cite{Tartakovsky2016}, the equation for the joint PDF $f_{\boldsymbol{X} \boldsymbol{Z}}(\boldsymbol{x}_\text{p},\boldsymbol{z};t)$ of the model input $\boldsymbol{Z}_i$ and output $\boldsymbol X_\text{p}(t)$ is exact:  
\begin{align}
    \frac{\partial f_{\boldsymbol{X} \boldsymbol{Z} }}{\partial t} + \nabla_{{\boldsymbol{x}}_\text{p}} \cdot \left[ \left(  \boldsymbol{u}_\text{p} + \sum_{i=1}^N \boldsymbol{z}_i \circ \boldsymbol{\varphi}_i(\boldsymbol{x}_\text{p},t)  \right) f_{\boldsymbol{X}\boldsymbol{Z}} \right] = 0.
    \label{eq: position_liouville_general}
\end{align}
This result can be derived via either the Liouville theorem~\cite{moyal1949stochastic,  soong1974random} or the method of distributions~\cite{Tartakovsky2016,dominguez2021lagrangian,dominguez2023MoC}.


Since $f_{\boldsymbol X}(\boldsymbol x_\text{p};t)$ is the marginal of $f_{\boldsymbol X \boldsymbol{Z}}(\boldsymbol x_\text{p},\boldsymbol{z};t)$, i.e.,
\[
f_{\boldsymbol X}(\boldsymbol x_\text{p};t) =  \int f_{\boldsymbol X \boldsymbol{Z}}(\boldsymbol x_\text{p},\boldsymbol{z};t) \text d \boldsymbol{z},
\]
the integration of~\eqref{eq: position_liouville_general} yields 
\begin{align}\label{eq:unclosed}
    \frac{\partial f_{\boldsymbol{X}}}{\partial t}+\nabla_{\boldsymbol{x}_\text{p}} \cdot \left(  \boldsymbol{u}_\text{p} f_{\boldsymbol{X}} \right) = - \nabla_{\boldsymbol{x}_\text{p}} \cdot \left( \int \sum_{i=1}^{N}\boldsymbol{z}_i \circ \boldsymbol{\varphi}_i f_{\boldsymbol{X}\boldsymbol{Z}} \, \text d \boldsymbol{z}  \right).
\end{align}
As mentioned above, the derivation of a PDF equation for $f_{\boldsymbol{X}}$ requires a closure to approximate the right hand side term. Comparison of~\eqref{eq:unclosed} and~\eqref{eq: position_fokker_general} indicates that the Langevin and Liouville approaches yield the identical predictions of the PDF $f_{\boldsymbol X}(\boldsymbol x_\text{p};t)$ if 
\begin{align}
     \nabla_{\boldsymbol{x}_\text{p}} \cdot \left( \boldsymbol{\mathcal D} \nabla_{\boldsymbol{x}_\text{p}} f_{\boldsymbol{X}}+ \int \sum_{i=1}^{N} \boldsymbol{z}_i \circ \boldsymbol{\varphi}_i f_{\boldsymbol{X}\boldsymbol{Z}} \, \text d \boldsymbol{z} \right) = 0.
    \label{eq: diffusion_tensor_back_fxp}
\end{align}
If a solution of a Fokker-Planck equation is known, the above identity may be used to develop an equivalent Liouville model. 
Taking moments of \eqref{eq: position_fokker_general} and~\eqref{eq: position_liouville_general}, similar relations can be found for each moment, adding constrains to the Liouville approach to match an equivalent Langevin model.
A moment model based on the Liouville approach is provided in Appendix~\ref{sec: moment_models}.

\subsection{Particle velocity}
\label{sec: PDF_models_velocity}

A class of models of particle-laden flows takes $\mathbf g$ in~\eqref{eq: tracer_deterministic_dxdt} to represent the drag force acting on a particle (see Ref.~\cite{mashayek2003analytical} and references therein). This force is proportional to the difference between the carrier flow velocity at the point $\boldsymbol x_\text{p}$ occupied by the particle, $\boldsymbol u(\boldsymbol x_\text{p},t)$, and the particle velocity, $\boldsymbol{u}_\text{p}(t)$, so that~\eqref{eq: tracer_deterministic_dxdt} is replaced with
\begin{align}\label{eq:velocity_deterministic}
    \frac{\text{d} \boldsymbol{x}_\text{p}}{\text{d}t} = \boldsymbol{u}_\text{p}, \qquad
    \frac{\text{d} \boldsymbol{u}_\text{p}}{\text{d}t} = \frac{\mathcal{F}}{\tau_\text{p}} \left(\boldsymbol{u} - \boldsymbol{u}_\text{p}\right).
\end{align}
The coefficient of proportionality comprises the correction factor $\mathcal{F}$, which accounts for a possible deviation from the Stokes drag (due to, e.g., finite particle-Reynolds number, particle Mach number, volume fraction of the particles), and the particle's response time $\tau_{\text{p}}$.

\subsubsection{Langevin approach}
\label{sec: PDF_models_velocity_Langevin}

A probabilistic treatment of~\eqref{eq:velocity_deterministic}, which is referred to velocity Langevin model~\cite{lattanzi2020stochastic}, uses Wiener increments to account for velocity fluctuations:
\begin{subequations}\label{eq: velocity_langevin}
\begin{align}
    \text{d}{\boldsymbol{X}_\text{p}} &= {\boldsymbol{U}_\text{p}} \text{d}t , 
    \label{eq: velocity_langevin_x} \\
    \text{d}{\boldsymbol{U}_\text{p}} &= \frac{\mathcal{F}}{\tau_\text{p}}\left[ \boldsymbol{u}({\boldsymbol{X}_\text{p}},t) - {\boldsymbol{U}_\text{p}} \right] \text{d}t + \boldsymbol{b}_u( {\boldsymbol{X}_\text{p}},{\boldsymbol{U}_\text{p}},t) \, \text{d}\boldsymbol{W}_u.
    \label{eq: velocity_langevin_u}
\end{align}
\end{subequations}
The Fokker-Planck equation for the joint PDF $f_{\boldsymbol{X}\boldsymbol{U}}\left( {\boldsymbol{x}_\text{p}},{\boldsymbol{u}_\text{p}};t \right)$ is given by 
\begin{align}
\begin{split}
    \frac{\partial f_{\boldsymbol{X} \boldsymbol{U}}}{\partial t} &+ \nabla_{\boldsymbol{x}_\text{p}} \cdot ( \boldsymbol{u}_\text{p} f_{\boldsymbol{X} \boldsymbol{U}} ) 
    + \nabla_{\boldsymbol{u}_\text{p}} \cdot \left[ \frac{\mathcal{F}}{\tau_\text{p}} ( \boldsymbol{u} -\boldsymbol{u}_\text{p} ) f_{\boldsymbol{X} \boldsymbol{U}}\right] = \nabla_{\boldsymbol{u}_\text{p}} \cdot ( \boldsymbol{\mathcal{D}} \nabla_{\boldsymbol{u}_\text{p}} f_{\boldsymbol{X}\boldsymbol{U}} ),
\end{split}
\label{eq: velocity_fokker_general}
\end{align}
with prescribed functions $\boldsymbol{u}=\boldsymbol{u}(\boldsymbol{x}_\text{p},t)$ and $\boldsymbol{\mathcal{D}}(\boldsymbol{x}_\text{p},\boldsymbol{u}_\text{p},t) =\boldsymbol{b}_u^{} \boldsymbol{b}_u^\top/2$.

\subsubsection{Liouville approach}
\label{sec: PDF_models_velocity_Liouville}

Our Liouville model relies on random variables $\boldsymbol{Z}_i$ and $\boldsymbol{\Xi}_i$, with PDFs $f_{\boldsymbol{Z}}(\boldsymbol{z})$ and $f_{\boldsymbol{\Xi}}(\boldsymbol{\xi})$, and basis functions given by $\boldsymbol{\phi}_i(\boldsymbol{X}_\text{p},\boldsymbol{U}_\text{p},t)$ and $\boldsymbol{\varphi}_i(\boldsymbol{X}_\text{p},\boldsymbol{U}_\text{p},t)$ to represent the stochastic particle dynamics:
\begin{subequations} \label{eq: velocity_ODE_liouville_general}
\begin{align}
    \frac{\text{d} \boldsymbol{X}_\text{p}}{\text{d}t} &= \boldsymbol{U}_\text{p} + \sum_{i=1}^{N}\boldsymbol{Z}_i \circ \boldsymbol{\varphi}_i(\boldsymbol{X}_\text{p},\boldsymbol{U}_\text{p},t), 
    \label{eq: velocity_ODE_liouville_general_X} \\ 
    \frac{\text{d} \boldsymbol{U}_\text{p}}{\text{d}t} &= \frac{\mathcal{F}}{\tau_\text{p}} \left[ \boldsymbol{u}({\boldsymbol{X}_\text{p}},t) - {\boldsymbol{U}_\text{p}} \right] + \sum_{i=1}^{N}\boldsymbol{\Xi}_i \circ \boldsymbol{\phi}_i(\boldsymbol{X}_\text{p},\boldsymbol{U}_\text{p},t).
    \label{eq: velocity_ODE_liouville_general_U}
\end{align}   
\end{subequations}
The corresponding Liouville equation for the joint PDF $ f_{\boldsymbol{X}\boldsymbol{U}\boldsymbol{Z} \boldsymbol{\Xi}}(\boldsymbol{x}_\text{p},\boldsymbol{u}_\text{p},\boldsymbol{z},\boldsymbol{\xi};t)$ is   
\begin{align}
\begin{split}
    \frac{\partial f_{\boldsymbol{X}\boldsymbol{U}\boldsymbol{Z} \boldsymbol{\Xi}}}{\partial t} &+ \nabla_{\boldsymbol{x}_\text{p}}\cdot \left[ \left( \boldsymbol{u}_\text{p}+ \sum_{i=1}^{N}\boldsymbol{z}_i \circ \boldsymbol{\varphi}_i(\boldsymbol{x}_\text{p},\boldsymbol{u}_\text{p},t)  \right) f_{\boldsymbol{X}\boldsymbol{U}\boldsymbol{Z} \boldsymbol{\Xi}} \right]  \\
    &+ \nabla_{\boldsymbol{u}_\text{p}}\cdot \left[ \left(\frac{\mathcal{F}}{\tau_\text{p}}\left( \boldsymbol{u}(\boldsymbol{x}_\text{p},t)-\boldsymbol{u}_\text{p} \right)  + \sum_{i=1}^{N} \boldsymbol{\xi}_i \circ \boldsymbol{\phi}_i(\boldsymbol{x}_\text{p},\boldsymbol{u}_\text{p},t) \right) f_{\boldsymbol{X}\boldsymbol{U}\boldsymbol{Z} \boldsymbol{\Xi}}\right] = 0.
\end{split}
\label{eq: velocity_Liouville_general}
\end{align}
Since $f_{\boldsymbol{X}\boldsymbol{U}}$ is the marginal of $f_{\boldsymbol{X}\boldsymbol{U}\boldsymbol{Z} \boldsymbol{\Xi}}$, the integration of~\eqref{eq: velocity_Liouville_general} over the coefficients yields an unclosed PDF equation for $f_{\boldsymbol{X}\boldsymbol{U}}$. Comparing that equation with the Fokker-Planck equation~\eqref{eq: velocity_fokker_general} yields a compatibility condition, 
\begin{align}
\begin{split}
 &\nabla_{\boldsymbol{x}_\text{p}} \cdot \left[  \int \sum_{i=1}^{N} \boldsymbol{z}_i  \circ \boldsymbol{\varphi}_i \left( \int f_{\boldsymbol{X}\boldsymbol{U}\boldsymbol{Z}\boldsymbol{\Xi}} \, \text d \boldsymbol{\xi} \right) \text{d} \boldsymbol{z} \right] + 
 \nabla_{\boldsymbol{u}_\text{p}} \cdot \left[ \boldsymbol{\mathcal{D}} \nabla_{\boldsymbol{u}_\text{p}} f_{\boldsymbol{X}\boldsymbol{U}} + \int \sum_{i=1}^{N} \boldsymbol{\xi}_i\circ \boldsymbol{\phi}_i \left(\int f_{\boldsymbol{X}\boldsymbol{U}\boldsymbol{Z}\boldsymbol{\Xi}} \, \text d \boldsymbol{z} \right) \text{d}\boldsymbol{\xi} \right] = 0,
\end{split}
\label{eq: diffusion_tensor_back}
\end{align}
for the Langevin and Liouville approaches. 
Relation~\eqref{eq: diffusion_tensor_back} can be used to compare both Fokker-Planck and Liouville approaches. We note that the Liouville approach has additional degrees as compared to the Fokker-Planck approach.
By comparing moments of both approaches, additional relations can be found.


\section{Analytical solutions to Liouville models} 
\label{sec: PDF_solutions}

\subsection{Particle trajectory}
\label{sec: PDF_solutions_position}

Lattanzi \textit{et al.}~\cite{lattanzi2020stochastic} considered a one-dimensional version of the Fokker-Planck equation~\eqref{eq: position_fokker_general} with constant velocity $u_\text{p}$ and diffusion tensor ${b_x}_{i j} = \sqrt{2 D} \delta_{i j}$,
\begin{align}
    \frac{\partial f_{X}}{\partial t} + u_\text{p} \frac{\partial f_X}{\partial x_\text{p}}= D \frac{\partial^2 f_{X}}{\partial x_\text{p}^2},
    \label{eq: position_fokker_1D}
\end{align}
subject to the deterministic initial condition, $f_{X}(x_\text{p};0)=\delta(x_\text{p})$. Its solution,
\begin{align}
    f_{X}(x_\text{p};t) = \frac{1}{\sqrt{4 \pi D t}} \exp \left( -\frac{\left( x_\text{p} -  u_\text{p} t \right)^2}{4 D t} \right),
    \label{eq: heat_kernel_solution}
\end{align} 
provides a probabilistic prediction of the particle trajectory, $X_\text{p}(t)$, within the Langevin framework. As expected, this solution is Gaussian.

In the corresponding one-dimensional Liouville approach, we search for relations to define the random forcings such that the solution of the Fokker-Planck equation and the Liouville equation coincide.
Starting from~\eqref{eq: position_fokker_1D} and~\eqref{eq: heat_kernel_solution}, we take the first moment of the Fokker-Plank solution and compare that with the first moment of the Liouville approach.
In particular, if we employ the law of the unconscious statistician (LOTUS)~\cite{flury2013first} to the characteristics of the Liouville equation~\eqref{eq: position_liouville_general},
\begin{align}
    \frac{\text{d} \hat{x}_\text{p}}{\text{d}t} = u_\text{p} + \sum_{i=1}^N z_i \varphi_i(t),
\end{align}
to find the first moment of the particle position which according to~\eqref{eq: heat_kernel_solution} is $\bar{X}_\text{p}=u_\text{p}t$, we find
\begin{align}
    \bar{X}_\text{p} = u_\text{p}t + \int \left( \sum_{i=1}^N z_i \int_{0}^t \varphi_i(t') \text{d} t' \right) f_{\boldsymbol{Z}}(\boldsymbol{z})\text{d}\boldsymbol{z},
    \label{eq: particle_position_sol_mon1_condition}
\end{align}
where we have applied the deterministic initial condition $X_\text{p}(0)=0$, according to the heat kernel solution~\eqref{eq: heat_kernel_solution}.
The second term in the right hand side of~\eqref{eq: particle_position_sol_mon1_condition} must be zero to match the solution of the Fokker-Planck equation.
We therefore find a condition to match the first moment 
\begin{align}
    \int \left( \sum_{i=1}^N z_i \int_{0}^t \varphi_i(t') \text{d} t' \right) f_{\boldsymbol{Z}}(\boldsymbol{z}) \text{d}\boldsymbol{z} = 0 .
    \label{eq: particle_position_sol_mon1_condition_II}
\end{align}
Then, looking at the second moment of the solution~\eqref{eq: heat_kernel_solution}, $\overline{{X_\text{p}^\prime}^2} = 2 D t $, using the same procedure we find the condition
\begin{align}
     \int [ u_\text{p}t + \sum_{i=1}^N {z}_i \int_{0}^t \varphi_i(t') \text{d} t' - \bar{X}_\text{p}^2 f_{\boldsymbol{Z}}(\boldsymbol{z})  \text{d}\boldsymbol{z} = 2 D t .
     \label{eq: particle_position_sol_mon2_condition_II}
\end{align}
Conditions~\eqref{eq: particle_position_sol_mon1_condition_II} and~\eqref{eq: particle_position_sol_mon2_condition_II} can be fulfilled by considering a single random coefficient $\Xi \equiv Z_i$, for $i=1,\dots, N$, which when used in~\eqref{eq: particle_position_sol_mon1_condition_II} leads to
\begin{align}
    \bar{\Xi}\sum_{i=1}^N \int_{0}^{t}\varphi_i(t^\prime)\text{d}t^\prime = 0, \qquad \bar{\Xi} = 0 .
\end{align}
Then, applying this to also~\eqref{eq: particle_position_sol_mon2_condition_II} gives
\begin{align}
    \sigma_{\Xi} \sum_{i=1}^N \int_{0}^{t}\varphi_i(t^\prime) \text{d}t^\prime = \sqrt{2Dt},
\end{align}
that can be fulfilled by considering a single function $\varphi \equiv \varphi_i$ such that 
\begin{align}
    \varphi = \sqrt{\frac{D}{2 t}} , \qquad \sigma_{\Xi} = 1.
\end{align}
Note that in general, basis function may be used to approximate the source by a finite number of terms $\sum_{i=1}^N \varphi_i(t) \approx \sqrt{D/(2t)}$.
The random source term can be simply described by a single term $\Xi \, \varphi(t)$ such that~\eqref{eq: position_liouville_general} reduces to
\begin{align}\label{eq:3.3}
    \frac{\partial f_{X \Xi}}{\partial t} + \left( u_\text{p} + \xi \sqrt{\frac{D}{2 t}} \right) \frac{\partial f_{X \Xi} }{ \partial x_\text{p}} = 0.
\end{align}
Since the initial position of the particle, $X_p(0) = 0$ is known with certainty, i.e., deterministic, this Liouville equation is subject to the initial condition $f_{X \Xi}(x_\text{p},\xi;0)=f_{\Xi}(\xi)\delta(x_\text{p})$.
Its solution, obtained via the method of characteristics in Appendix~\ref{app:derivations}, is
\begin{align}
    f_{X \Xi}(x_\text{p},\xi;t) = f_\Xi(\xi) \delta(
    x_\text{p} - u_\text{p} t - \xi\sqrt{2 D t}
    ).
    \label{eq: position_joint_aXp_1}
\end{align}
The PDF of the particle position, $f_{X}(x_\text{p};t)$, is the marginal of $f_{X \Xi}$ computed by integrating~\eqref{eq: position_joint_aXp_1} over $\xi$, while accounting for the properties of the the Dirac delta function $\delta(-x)=\delta(x)$ and $|\gamma|\delta(\gamma x)=\delta(x)$ for a given scalar $\gamma$,
\begin{align}
\begin{split}
    f_{X}(x_\text{p};t) 
    = \frac{1}{\sqrt{2 D t}} f_{\Xi}\left(  \frac{x_\text{p}-u_\text{p} t}{\sqrt{2 D t}} \right).
    \label{eq: heat_kernel_solution_general}
\end{split}
\end{align}

\begin{figure}[hbt!]
    \centering
    \includegraphics[width=0.4\textwidth]{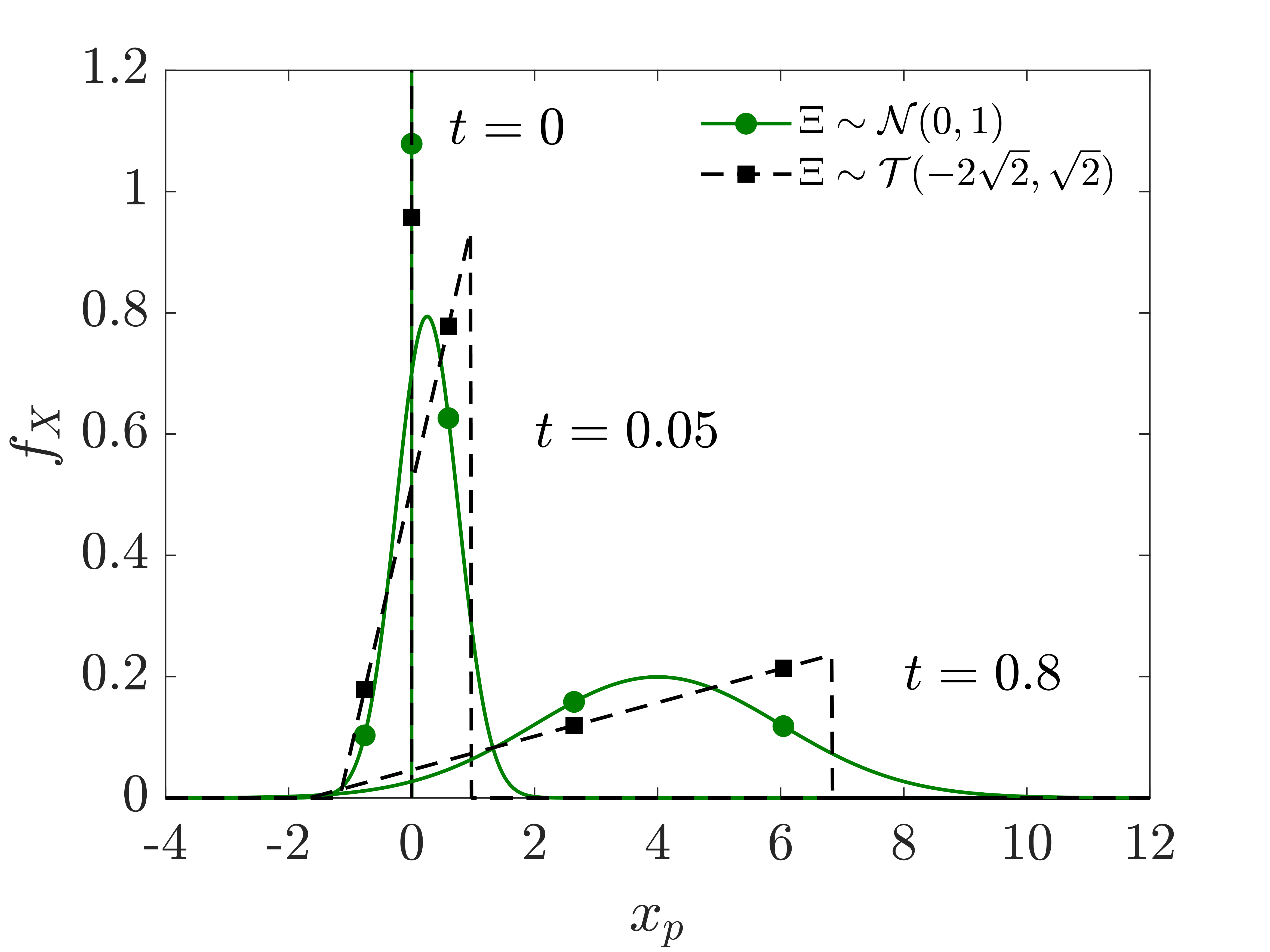}
    \caption[]{Particle position PDF $f_X$ at times $t=[0,\ 0.05, \ 0.8]$ given by~\eqref{eq: heat_kernel_solution_general} with $D=2.5$ and $u_\text{p}=5$ for normal and triangular distributions of the parameter $\Xi$. The uniform distribution has been omitted for clarity (see Fig.~\ref{fig: PL_fx_unif}).}
    \label{fig: fX_atTimes}
\end{figure}

\begin{figure*}[t]
    \centering
    \subfloat[]{
    \label{fig: PL_fx_norm}
    \includegraphics[width=0.33\textwidth]{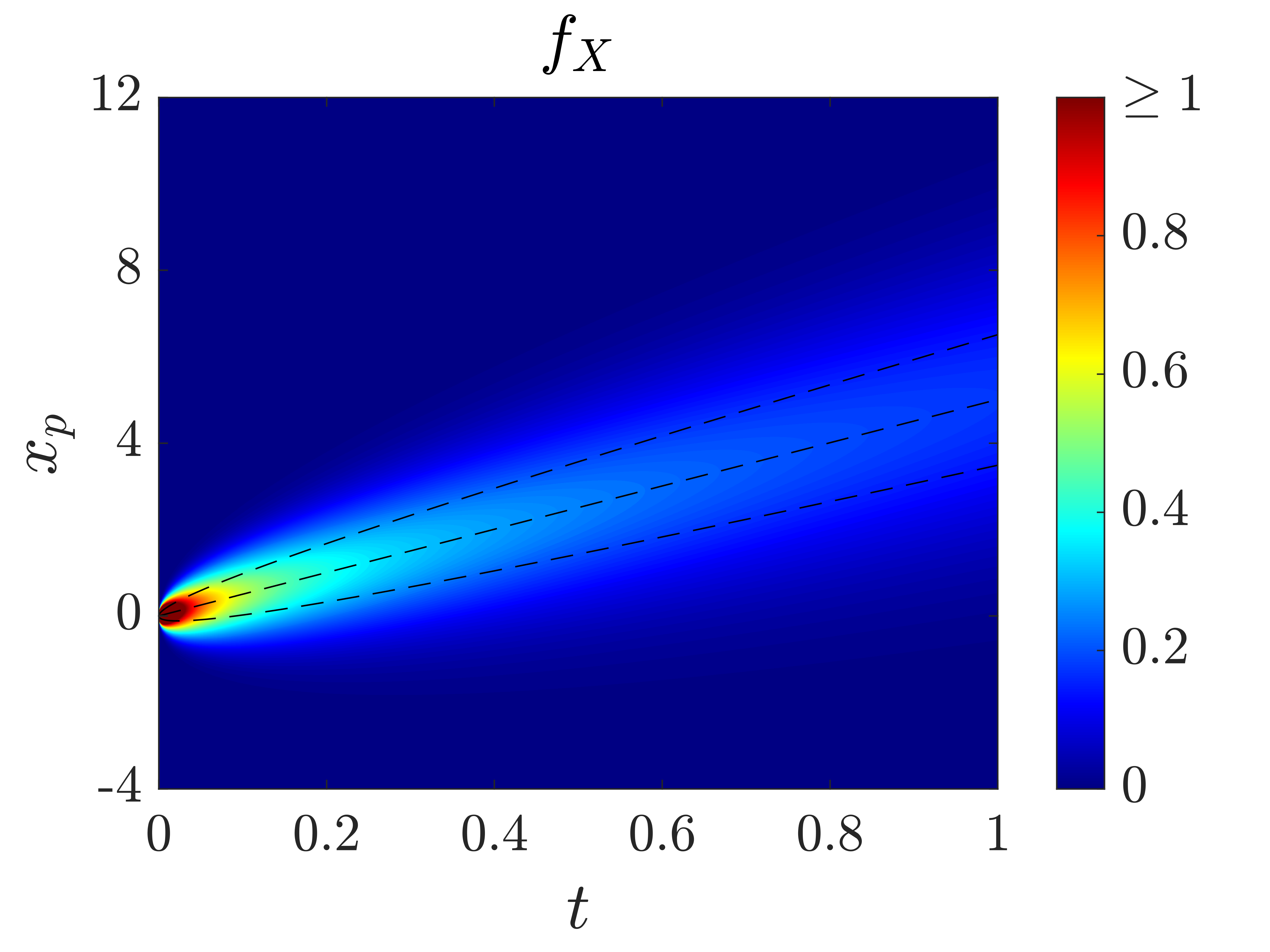}} 
    \subfloat[]{
    \label{fig: PL_fx_unif}
    \includegraphics[width=0.33\textwidth]{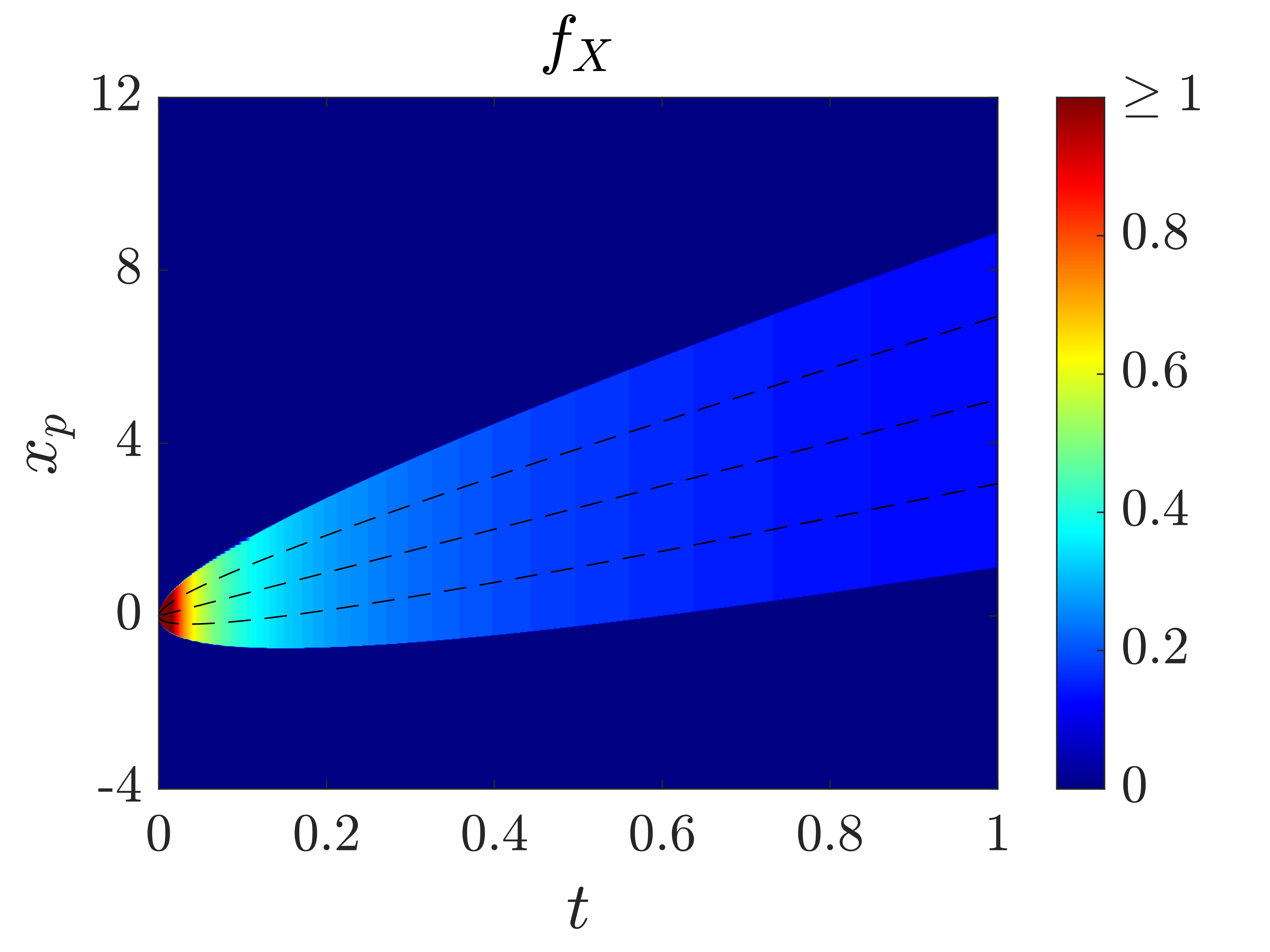}} 
    \subfloat[]{
    \label{fig: PL_fx_tria}
    \includegraphics[width=0.33\textwidth]{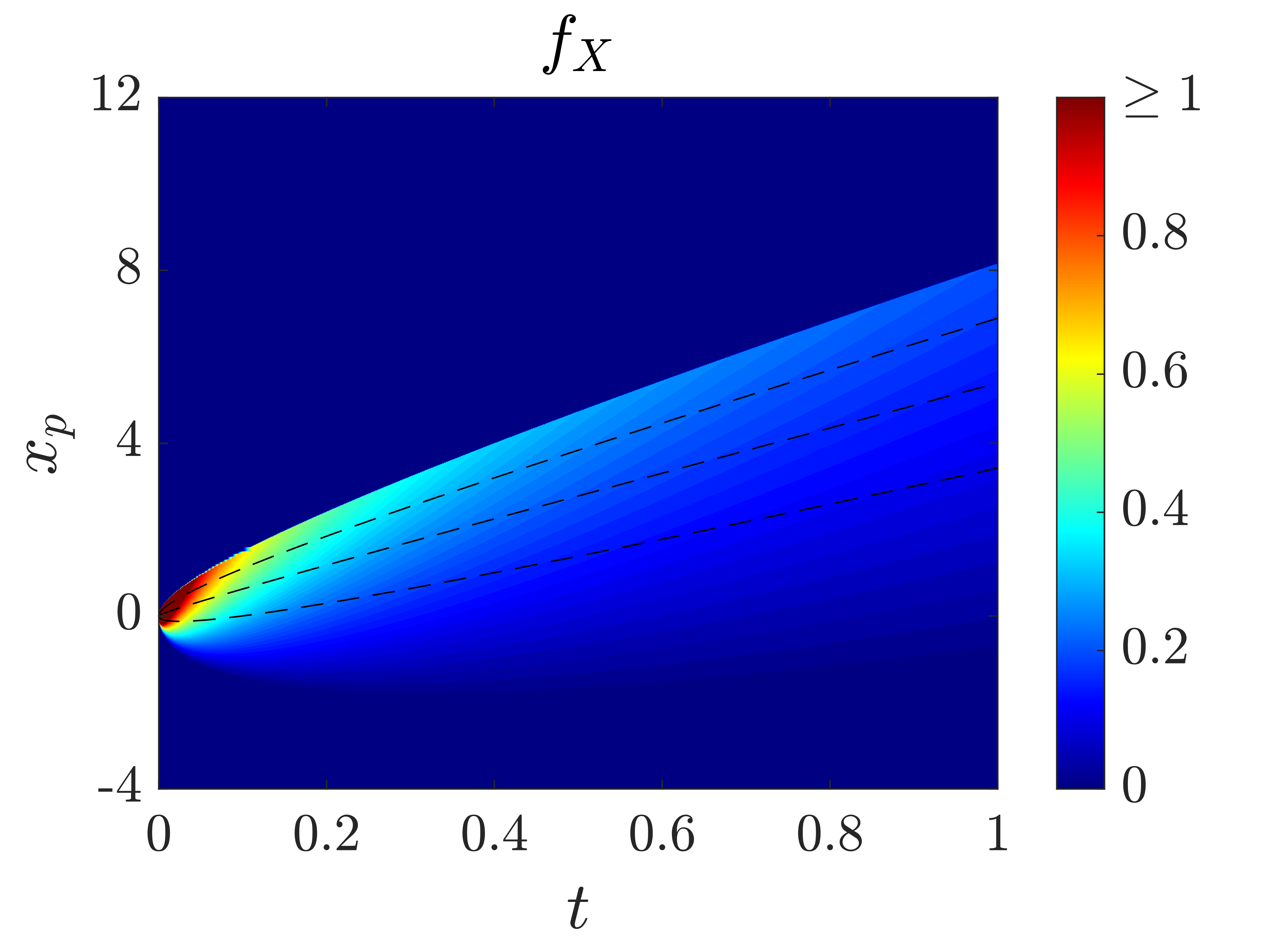}} 
    \caption[]{Temporal evolution of $f_X(x_\text{p};t)$, PDF of the particle trajectory $X_\text{p}(t)$, predicted by the Liouville solution~\eqref{eq: heat_kernel_solution_general} for (a) normal, (b) uniform and (c) triangular distributions of the parameter $\Xi$. The deterministic parameters are set to $D=2.5$ and $u_\text{p}=5$. 
    }
    \label{fig: PL_fx}
\end{figure*}

If $f_\Xi(\xi)$ is standard normal, $\Xi \sim \mathcal{N}(0,1)$, then the Liouville solution~\eqref{eq: heat_kernel_solution_general} reduces to the Gaussian solution~\eqref{eq: heat_kernel_solution} predicted by the Langevin approach. Otherwise, the PDF $f_X$ in~\eqref{eq: heat_kernel_solution_general} is non-Gaussian (Fig.~\ref{fig: fX_atTimes}). 
The temporal evolution of the particle-trajectory PDF $f_X(x_\text{p};t)$ is depicted in Figure~\ref{fig: PL_fx} for the normal, $\mathcal{N}(0,1)$; uniform, $\mathcal{U}(-\sqrt{3},\sqrt{3})$; and triangular, $\mathcal{T}(-2\sqrt{2},\sqrt{2})$; distributions of the parameter $\Xi$. These three distributions have the same mean, 0, and variance, 1, but different higher moments. 
The normal distribution gives rise to the Gaussian $f_X(x_\text{p};t)$, associated with the Wiener process. On the other hand, the triangular distribution translates into a skewed (non-Gaussian) PDF of the particle trajectory.
The uniform distribution provides the same statistics than the normal up to the second moments, despite both PDFs differ at any given time (Figs.~\ref{fig: PL_fx_norm} and~\ref{fig: PL_fx_unif}).


The compatibility condition~\eqref{eq: diffusion_tensor_back_fxp} may be employed to obtain the diffusion tensor $\mathcal{D}$ for the Fokker-Planck equation from the results obtained with the Liouville approach. 
In fact, the substitution of~\eqref{eq: position_joint_aXp_1} and~\eqref{eq: heat_kernel_solution_general} into~\eqref{eq: position_fokker_1D} proves the identity, showing consistency between the Langevin and Liouville approaches. 

\subsection{Particle velocity} 
\label{sec: PDF_solutions_velocity}

The so-called velocity Langevin model~\cite{lattanzi2020stochastic} deals with a one-dimensional version of
~\eqref{eq: velocity_langevin}, 
\begin{align}\label{eq: SOL_velocity_langevin}
    \text{d}{{X}_\text{p}} =  \tau_\text{p} {{U}_\text{p}} \text{d}t, \qquad
    \text{d}{{U}_\text{p}} = - {{U}_\text{p}}\text{d}t + \sqrt{2 D} \text{d}{W},
\end{align}
in a reference frame moving at the average particle velocity~\cite{lattanzi2020stochastic}.
The corresponding Fokker-Planck equation for the joint PDF $f_{X U}(x_\text{p},u_\text{p};t)$ of the particle trajectory $X_\text{p}(t)$ and velocity $U_\text{p}(t)$ is 
\begin{align}
    \frac{\partial f_{X U}}{\partial t} + \tau_\text{p} {u}_\text{p} \frac{\partial f_{X U}}{\partial x_\text{p}} - \frac{\partial }{\partial u_\text{p}} \left( u_\text{p} f_{X U}\right) = D \frac{\partial^2 f_{X U}}{\partial u_\text{p}^2}.
    \label{eq: SOL_velocity_fokker_fxpup}
\end{align}
If the initial particle position and velocity are known with certainty, $X_\text{p}(0) = 0$ and $U_\text{p}(0) = v_0$, then~\eqref{eq: SOL_velocity_fokker_fxpup} is subject to the initial condition $f_{X U}(x_\text{p}, u_\text{p};0)=\delta(x_\text{p}) \delta(u_\text{p}-v_0)$. Integration of this equation over  $x_\text{p}$ yields the Fokker-Planck equation for the marginal $f_{U}(u_\text{p};t)$.
\begin{align}
    \frac{\partial f_{U}}{\partial t}  - \frac{\partial }{\partial u_\text{p}}  \left( u_\text{p} f_{U}\right)   = D \frac{\partial^2  f_{U}}{\partial u_\text{p}^2},
    \label{eq: velocity_fokker_marginal_fu}
\end{align}
subject to the initial condition $f_{U}(u_\text{p};0)=\delta(u_\text{p}-v_0)$. This problem admits the Gaussian solution~\cite{risken1996fokker, pope2000turbulent, gardiner2009stochastic, lattanzi2020stochastic}
\begin{align}
    f_{U}(u_\text{p};t) = \frac{1}{\sqrt{2 \pi D ( 1 - \text{e}^{-2 t} ) }} \exp\left[-\frac{(u_\text{p} - v_0 \text{e}^{-t}  )^2 }{2 D ( 1 - \text{e}^{-2 t} ) } \right].
    \label{eq: velocity_solution_fu}
\end{align}

Our Liouville approach replaces the Langevin model~\eqref{eq: SOL_velocity_langevin} with the one-dimensional version of~\eqref{eq: velocity_ODE_liouville_general}. 
To find the random forcings that make coincide both approaches, we proceed similar as in the previous case (see Section~\ref{sec: PDF_solutions_position}).
Equivalently to the use of the LOTUS with the characteristics of the Liouville equation to obtain moments, the model developed in Appendix~\ref{sec: moment_models}, can also be used to find expressions for the moments using the Liouville approach.
In essence, we compare the moments of the solution of the Fokker-Planck equation~\eqref{eq: SOL_velocity_fokker_fxpup} to find the random forcings in the Liouville approach.
As in the previous example, we assume a single random coefficient can be used $\Xi\equiv \Xi_i=Z_i$ and the basis functions reduce to a deterministic function of time, $\varphi(t)$ and $\phi(t)$ for the position and velocity equations respectively.
Then, using the moment equations~\eqref{eq: velocity_moments} derived in Appendix~\ref{sec: velocity_moments}, one can find the closed-form functions
\begin{subequations} \label{eq: Velocity_phi_varphi}
\begin{align}
    \begin{split}
    \varphi =  \frac{\text{d} \sigma_X}{\text{d}t} -\tau_p \sigma_{U}  
    = \tau_\text{p} \sqrt{ D} \left(  \frac{1-2 \text{e}^{-t} + \text{e}^{-2t}}{\sqrt{2t-3+4 \text{e}^{-t} - \text{e}^{-2t}}}  -\sqrt{1 - \text{e}^{-2t}} \right),     
    \end{split} \\
    \phi &= \frac{\text{d}\sigma_U}{\text{d}t} + \sigma_U = \sqrt{\frac{D}{1-\text{e}^{-2t}  }}, 
\end{align}
\end{subequations}
where we have taken as in the previous example, $\bar{\Xi}=0$, and $\sigma_{\Xi}=1$.
The functions $\sigma_X(t)$ and $\sigma_U(t)$ are the standard deviations of $X_\text{p}(t)$ and $U_\text{p}(t)$ according to the Fokker-Planck equation~\eqref{eq: SOL_velocity_fokker_fxpup}, given by\cite{lattanzi2020stochastic}
\begin{subequations}\label{eq: velocity_sigmas}
\begin{align}
    \sigma_X &= \sqrt{\tau_\text{p}^2 D\left( 2t-3+4 \text{e}^{-t} - \text{e}^{-2t} \right)}, \\
    \sigma_U &= \sqrt{D\left( 1 - \text{e}^{-2t}\right)}.
\end{align}
\end{subequations}

In general, data-driven techniques can be used to adjust the random forcings in~\eqref{eq:intro_ODEs_random} by defining an optimization problem. 
See for example Ref.~\cite{dominguez2022adjoint} where the PDF of the random coefficients are learned from DNS with an adjoint-based data assimilation procedure using Fourier series as basis functions. 

\begin{figure*}[t]
	\centering
	\subfloat[]{
		\label{fig: VL_fx_norm}
		\includegraphics[width=0.33\textwidth]{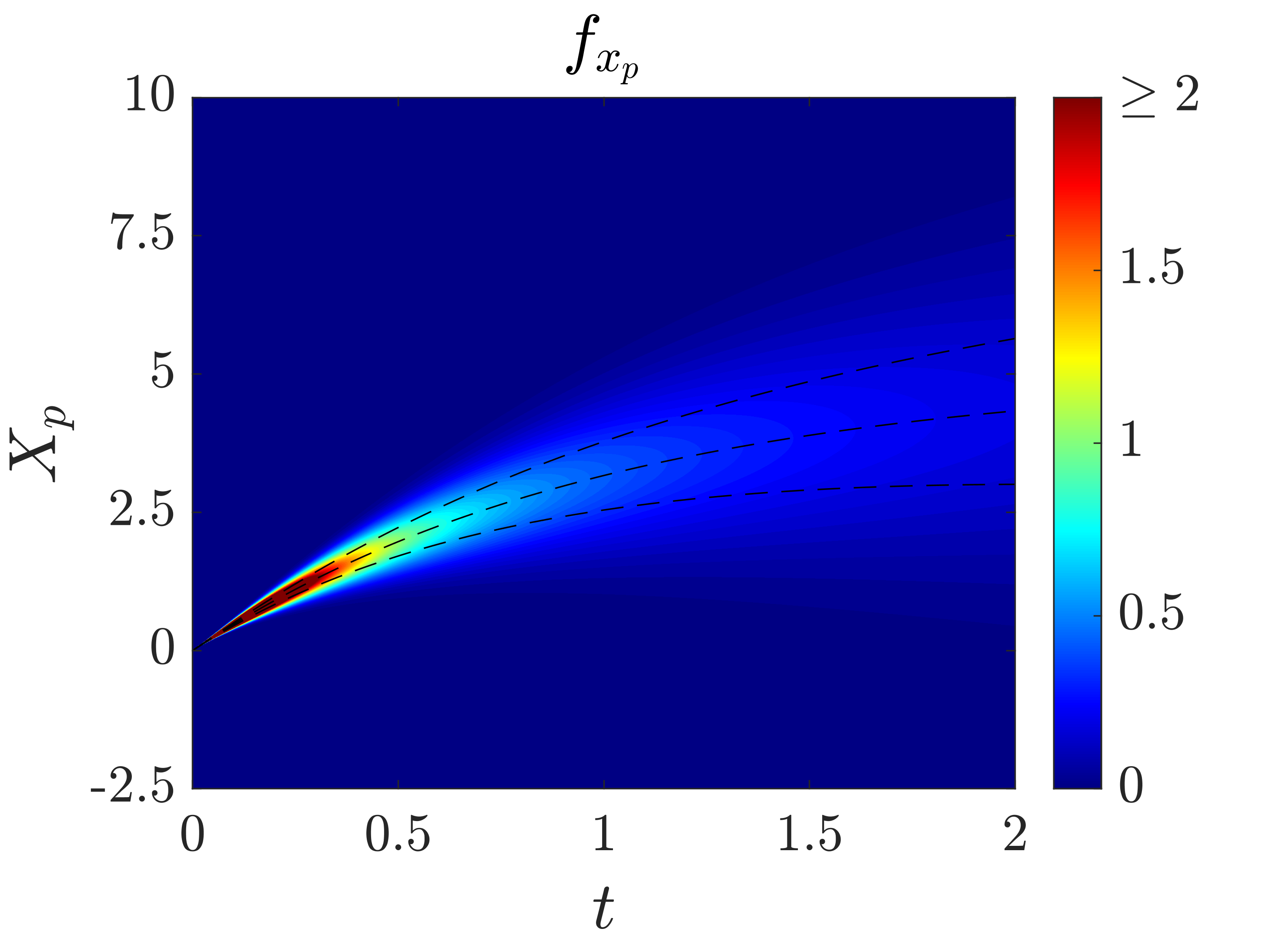}}
    \subfloat[]{
		\label{fig: VL_fx_unif}
		\includegraphics[width=0.33\textwidth]{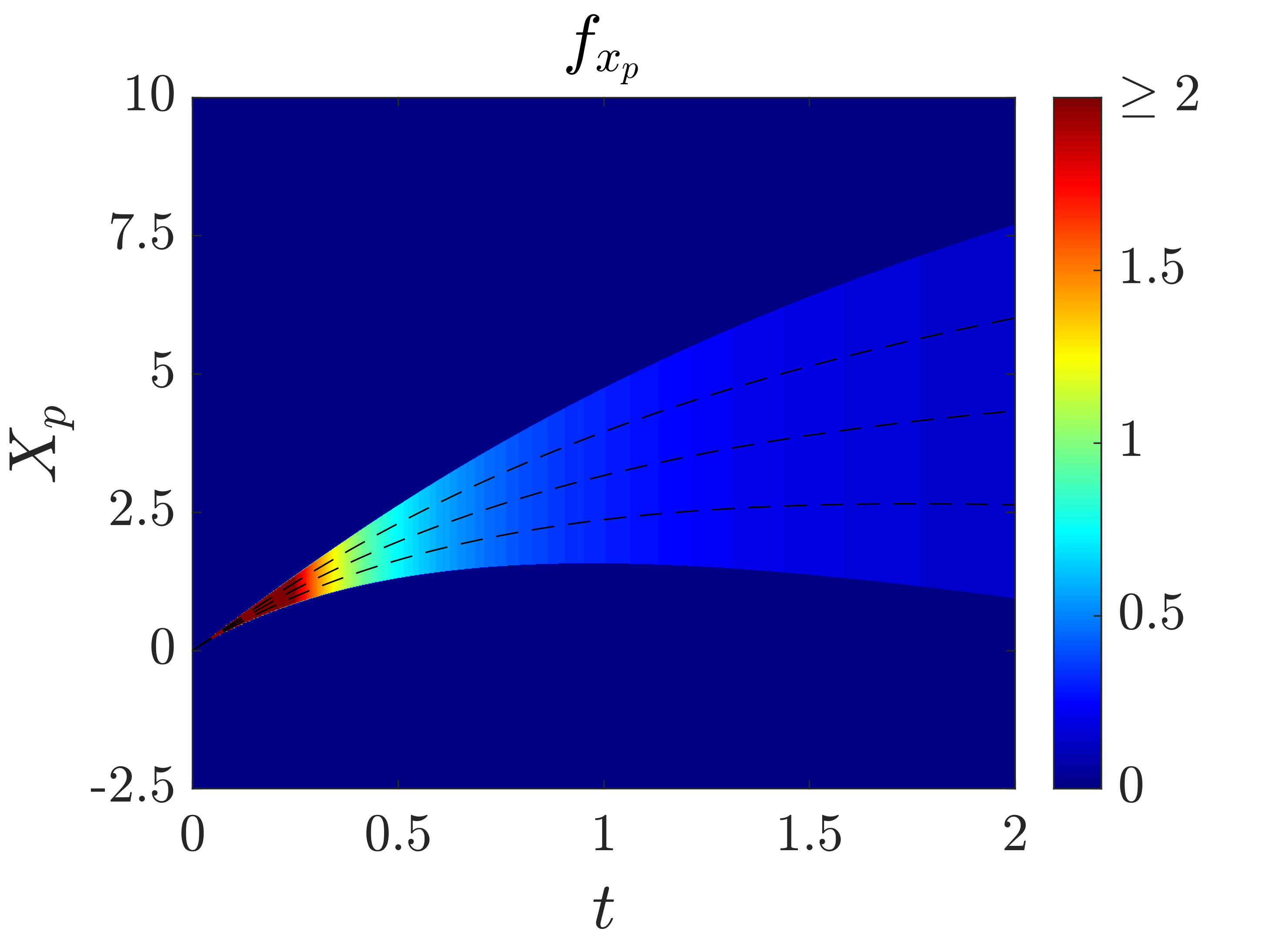}}
	\subfloat[]{
		\label{fig: VL_fx_tria}
		\includegraphics[width=0.33\textwidth]{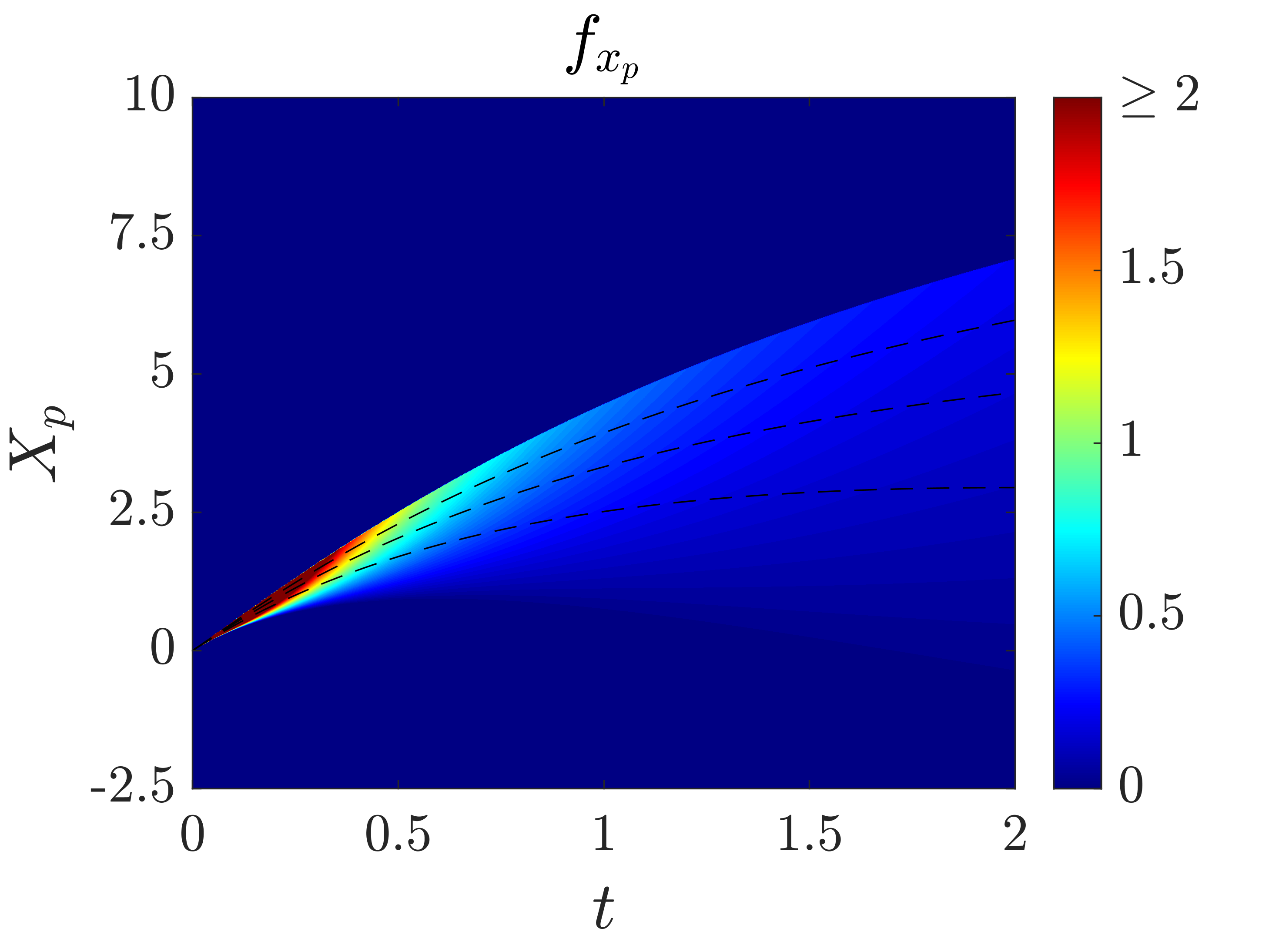}} \\
		\subfloat[]{
		\label{fig: VL_fu_norm}
		\includegraphics[width=0.33\textwidth]{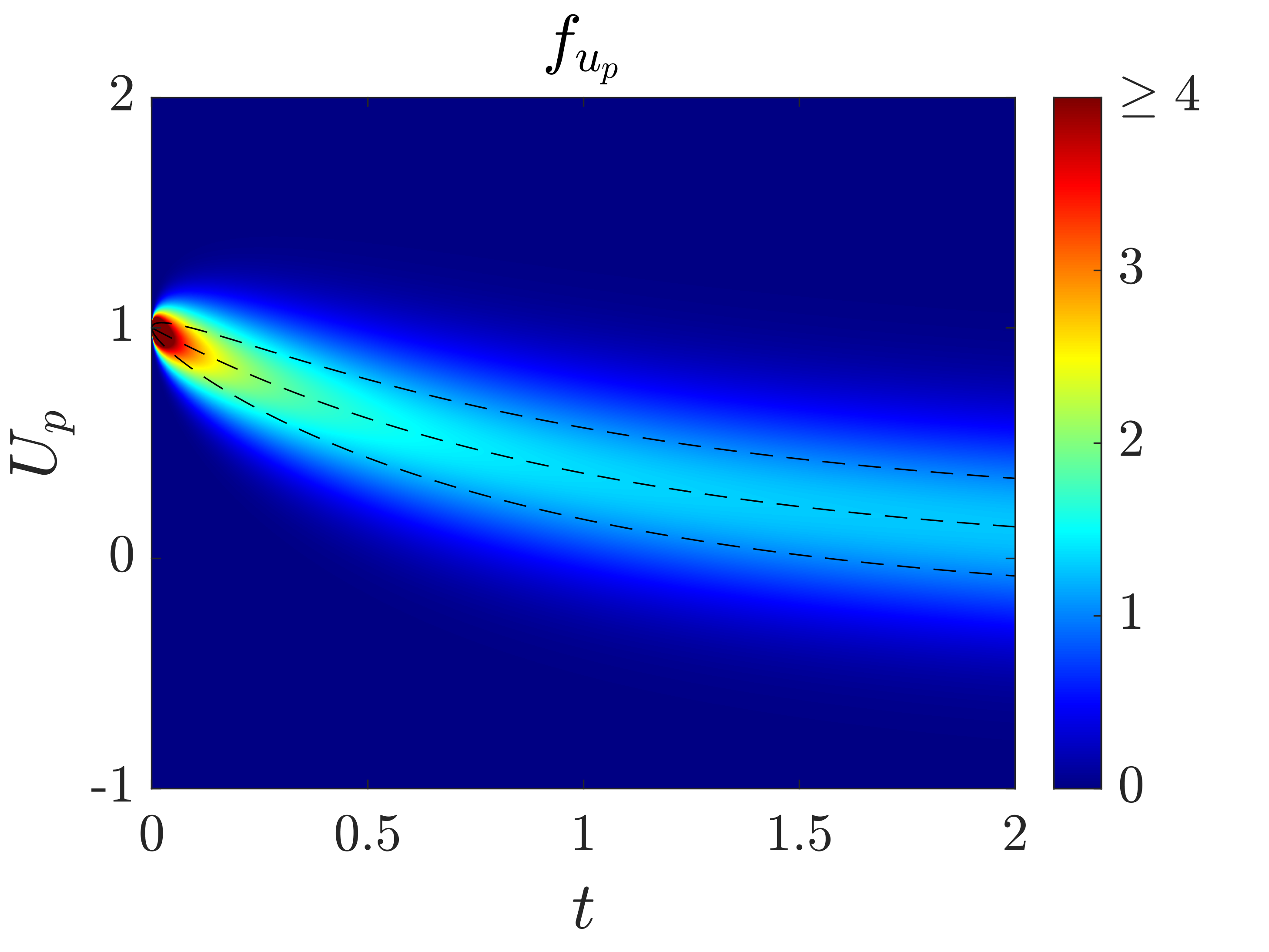}}
    \subfloat[]{
		\label{fig: VL_fu_unif}
		\includegraphics[width=0.33\textwidth]{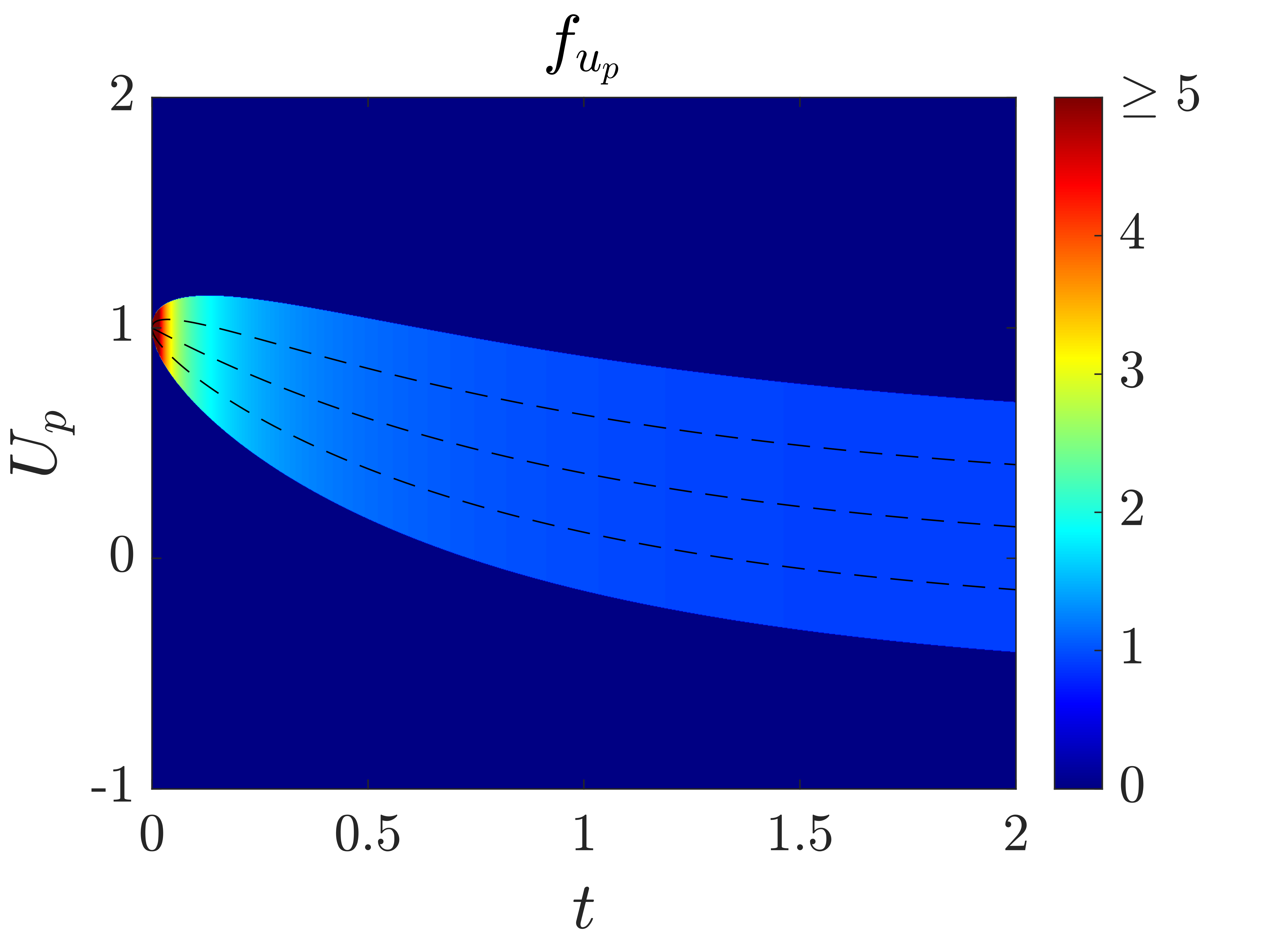}}
	\subfloat[]{
		\label{fig: VL_fu_tria}
		\includegraphics[width=0.33\textwidth]{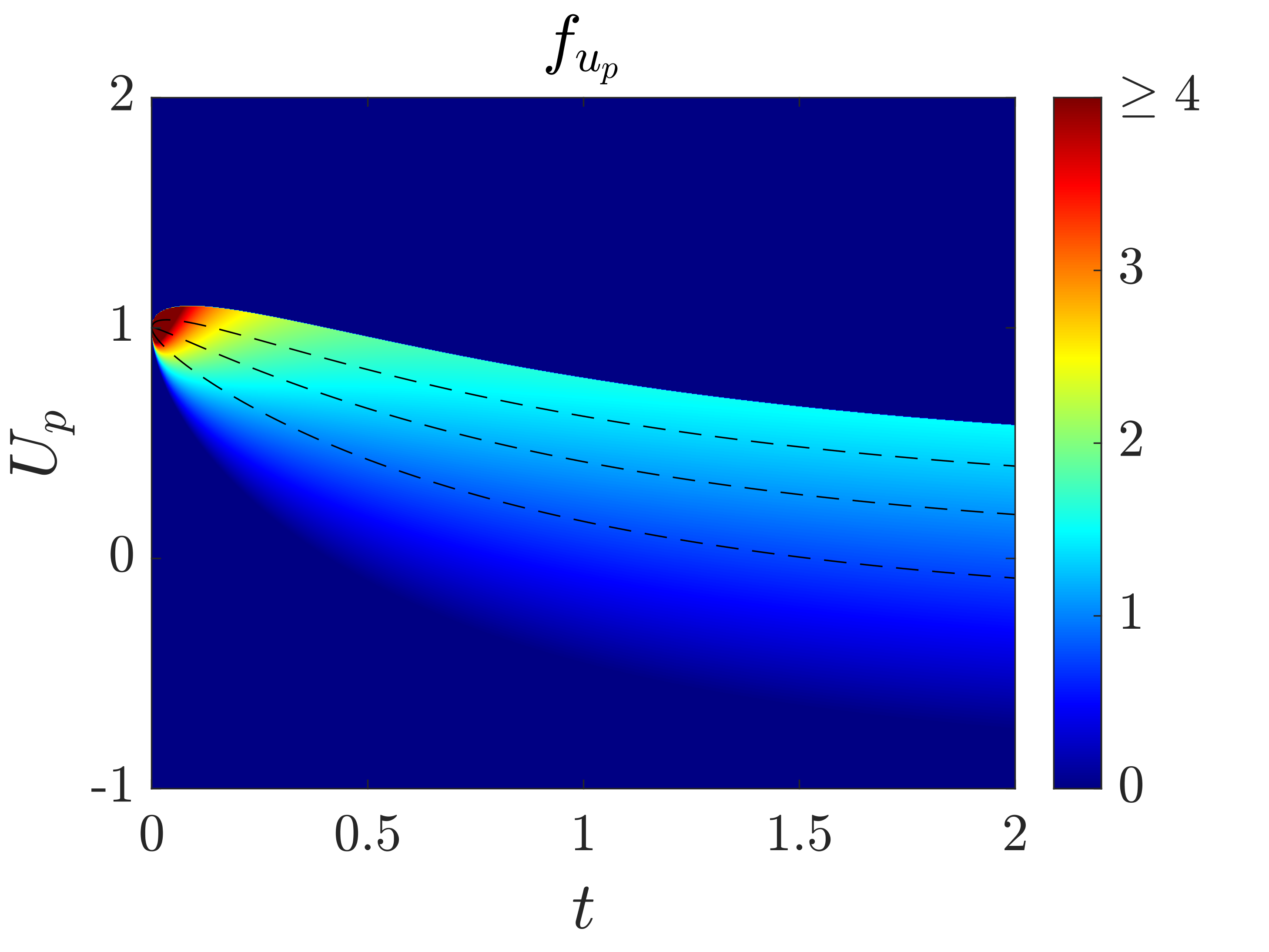}} 
	\caption[]{Temporal evolution of $f_X(x_\text{p};t)$ and $f_U(u_\text{p};t)$, PDFs of the particle trajectory $X_\text{p}(t)$ and velocity $U_\text{p}(t)$, predicted by the Liouville solutions~\eqref{eq: velocity_solution_general_f_xp} and~\eqref{eq: velocity_solution_general_f_up} for normal (left column), uniform (middle column) and triangular (right column) distributions of the parameter $\Xi$. The deterministic parameters are set to $\tau_\text{p}=10$, $v_0=1$ and $D = 1/50$.}
	\label{fig: VL_fu}
\end{figure*}

With the random forcings defined by the random coefficient $\Xi$ and the functions in~\eqref{eq: Velocity_phi_varphi}, our Liouville approach replaces the Langevin model ~\eqref{eq: SOL_velocity_langevin} with
\begin{align}
    \frac{\text{d} {X}_\text{p}}{\text{d}t} =  \tau_\text{p} U_\text{p} + \Xi \varphi, \qquad 
    \frac{\text{d} {U}_\text{p}}{\text{d}t} = - U_\text{p} + \Xi \phi,
    \label{eq: velocity_ODE_liouville_sol}
\end{align}
The Liouville equation for the joint PDF $f_{X U \Xi}(x_\text{p},u_\text{p},\xi;t)$ is 
\begin{align}
\begin{split}
    \frac{\partial f_{X U \Xi}}{\partial t} &+ \frac{\partial }{\partial x_\text{p}} \left[ \left(\tau_\text{p} u_\text{p}+\xi \varphi  \right) f_{X U \Xi} \right] 
    + \frac{\partial }{\partial u_\text{p}} \left[ \left(- u_\text{p} + \xi\phi \right) f_{X U \Xi}\right] = 0,
    \end{split}
    \label{eq: velocity_Liouville_3D}
\end{align}
subject to the initial condition $f_{X U \Xi}(x_\text{p},u_\text{p},\xi;0) = \delta(x_\text{p}) \delta(u_\text{p}-v_0) f_\Xi(\xi)$,  
which gives rise to the solution (Appendix~\ref{app:derivations}) 
\begin{align}
    \begin{split}
    f_{X U \Xi} = 
    \text{e}^{t}f_\Xi(\xi) \delta [  x_\text{p} - \tau_\text{p} v_0 ( 1 - \text{e}^{-t} ) - \xi\sigma_X ] \, \delta [ \text{e}^{t} (u_\text{p} - \xi \sigma_U ) - v_0 ].
    \end{split}
    \label{eq: velocity_solution_general_f_axpup}
\end{align}
The marginals of~\eqref{eq: velocity_solution_general_f_axpup} are the PDFs of the particle position and velocity,
%
\begin{align}
    f_{X}(x_\text{p};t) =   \frac{1}{\sigma_X} f_\Xi \left( \frac{x_\text{p} -\bar{X}_\text{p}}{\sigma_X} \right) ,
    \label{eq: velocity_solution_general_f_xp}
\end{align}
and
\begin{align}
    f_{U}(u_\text{p};t) &=  \frac{1}{\sigma_U}f_\Xi \left( \frac{u_\text{p}- \bar{U}_\text{p}}{\sigma_U} \right),
    \label{eq: velocity_solution_general_f_up}
\end{align}
where $\bar{X}_\text{p}(t)$ and $\bar{U}_\text{p}(t)$ are the means of $X_\text{p}(t)$ and $U_\text{p}(t)$, respectively (given in~\eqref{eq: velocity_SOLmoments}). 
Solutions~\eqref{eq: velocity_solution_general_f_xp} and~\eqref{eq: velocity_solution_general_f_up} are generalizations of the velocity Langevin model~\cite{lattanzi2020stochastic}, with~\eqref{eq: velocity_solution_general_f_up} reducing to~\eqref{eq: velocity_solution_fu} when the random variable $\Xi$ is standard normal, $\Xi \sim \mathcal{N}(0,1)$. 
The temporal evolution of the PDFs $f_X(x_\text{p};t)$ and $f_U(u_\text{p};t)$ is depicted in Figure~\ref{fig: VL_fu}, for the normal, uniform and triangular distributions of $\Xi$. 
The spatial variations of the PDFs of particle positions and velocity related to non-Gaussian effects (Figs.~\ref{fig: VL_fx_unif}--~\ref{fig: VL_fx_tria} and~\ref{fig: VL_fu_unif}--~\ref{fig: VL_fu_tria}) can not be represented by the Fokker-Plank model~\eqref{eq: SOL_velocity_fokker_fxpup} (Figs.~\ref{fig: VL_fx_norm} and~\ref{fig: VL_fu_norm}).
The Liouville approach may be used then to formulate non-Gaussian stochastic models.

When applied to~\eqref{eq: velocity_Liouville_3D}, the method of characteristics provides a geometric  interpretation of the dynamics of the joint PDF $f_{X U \Xi}(x_\text{p},u_\text{p},\xi;t)$ and its marginals. 
The joint PDF $f_{X U \Xi}$ is transported along the characteristics $\boldsymbol y(t) = [\hat x_\text{p}(t), \hat u_\text{p}(t), \xi]^\top$, defined by~\eqref{eq: velocity_characteristics_solution}. This process defines a flow map~\cite{dominguez2023MoC} and is akin to mapping the initial state $f_{X U \Xi}(x_\text{p},u_\text{p},\xi; 0)$ onto its counterpart at time $t$ via the method of transformations~\cite{soong1974random}: 
\begin{align}
    f_{X U \Xi}(x_\text{p},u_\text{p},\xi;0) = J(t) f_{X U \Xi}(x_\text{p},u_\text{p},\xi;t), \quad J = \left|\frac{\partial \, {\boldsymbol y}(t)}{ \partial \, {\boldsymbol y}(0)} \right|,
    \label{eq: velocity_MoT}
\end{align}
where $J(t)$ the determinant of the transformation Jacobian. Its direct evaluation from the characteristics equation~\eqref{eq: velocity_characteristics_solution_Up} yields $J(t) = \exp(-t)$, which is consistent with~\eqref{eq: velocity_solution_general_f_axpup} and~\eqref{eq: velocity_characteristics_solution_f}. 
Numerical evaluation of the Jacobian and the use of the method of transformations, i.e., using~\eqref{eq: velocity_MoT}, provides an alternative to computing the solution of~\eqref{eq: velocity_characteristics_solution_f}. 
\section{A Liouville model for fluidized homogeneous heating systems} 
\label{sec: FHHS}

In a fluidized homogeneous heating system (FHHS), particles with diameter $d_\text{p}$ are released into a current of carrier-fluid (with density $\rho_\text{f}$ and viscosity $\mu_\text{f}$). 
As time increases, the particle velocity $\boldsymbol U_\text{p}(t)$ deviates from the fluid velocity in such a way that the variances of the particle velocity along each component reach a steady value. The quantities  
\begin{align}
    T = \frac{1}{3} \overline{\boldsymbol{U}_\text{p}^\prime \cdot \boldsymbol{U}_\text{p}^\prime } \quad \text{and}\quad
    \Rey_T = \frac{\rho_\text{f} \, d_\text{p} }{\mu_\text{f} } \sqrt{T}
    \label{eq: FHHS_T_definition}
\end{align}
are referred to as the granular temperature and the thermal Reynolds number, respectively~\cite{lattanzi2022stochastic}.  Notice that a Reynolds decomposition is used for $\boldsymbol{U}_\text{p}=\bar{\boldsymbol{U}}_\text{p} + \boldsymbol{U}_\text{p}^\prime$.
They have been studied analytically via kinetic theory \cite{koch1999particle} or empirical treatment of Langevin models \cite{lattanzi2022fluid} and numerically via
PR-DNS \cite{tenneti2016stochastic} or Langevin equations \cite{lattanzi2022stochastic}. We use these results to ascertain the veracity of a new model based on our Liouville approach. 


In a reference frame moving with the average particle velocity~\cite{lattanzi2020stochastic,lattanzi2022stochastic}, a one-dimensional Liouville model of the random particle velocity $U_\text{p}(t)$ is 
\begin{align}
    \frac{\text{d} U_\text{p}}{\text{d}t} &= - \frac{1}{\tau_\text{p}} U_\text{p}  + \Xi \phi,
    \label{eq: FHHS_dudt_1}
\end{align}
such that $T=\sigma_{U}^2/3$.
The random variable $\Xi$ has zero mean, $\bar{\Xi}=0$, and standard deviation $\sigma_{\Xi} = \sigma_{\Xi}(\Rey_\text{m})$, which depends on the mean Reynolds number in the PR-DNS~\cite{lattanzi2022stochastic},
\begin{align*}
    \Rey_\text{m} = (1-\omega) \frac{\rho_\text{f} d_\text{p} }{\mu_\text{f} } |\boldsymbol{V}|,
\end{align*}
where $\boldsymbol{V}$ is the average relative velocity, and $\omega$ is the average particle solid volume fraction. 
%
The deterministic function,
%
\begin{align}
    \phi = \frac{1}{\tau_\text{p}} \left(1 - \text{e}^{-\mathcal{C}_1 t / \tau_\text{p}}\right)^{\mathcal{C}_2-1} \left[ 1 + \left( \mathcal{C}_1 \mathcal{C}_2 - 1 \right)\text{e}^{-\mathcal{C}_1 t / \tau_\text{p}} \right]
    \label{eq: FHHS_varphi}
\end{align}
is chosen phenomenologically, 
with the constants $\mathcal{C}_1$ and $\mathcal{C}_2$ characterizing the behavior of the granular temperature $T$ at early and late times, respectively. 
Both constants depend on $\Rey_\text{m}$, $\rho_\text{p}/\rho_\text{f}$ and $\omega$, and are adjusted to match the PR-DNS results \cite{tenneti2016stochastic, lattanzi2022fluid, lattanzi2022stochastic}. 

The Liouville equation for the joint PDF $f_{U \Xi }(u_\text{p},\xi;t)$ is
\begin{align}
    \frac{\partial f_{U\Xi}}{\partial t} + \frac{\partial }{\partial u_\text{p}}\left[ \left( -\frac{1}{\tau_\text{p}}u_\text{p} +\xi\phi  \right) f_{U\Xi}  \right] = 0.
    \label{eq: FHHS_Liouville_1}
\end{align}
Its solution is (Appendix~\ref{app:derivations})
\begin{subequations}\label{eq:sol3}
\begin{align}
\begin{split}
    f_{U\Xi} = f_{U\Xi}^0 \text{e}^{t / \tau_\text{p}} = \text{e}^{t / \tau_\text{p}} f_{\Xi}(\xi) \delta [ ( u_\text{p} - \xi \eta ) \text{e}^{t / \tau_\text{p}} ] 
    = \frac{1}{\eta}f_{\Xi}(\xi)\delta\left( \frac{u_\text{p}}{\eta} - \xi \right) ,
\end{split}\label{eq:FHHS_jointsol}
\end{align}
where
\begin{align}
\eta(t) = ( 1 - \text{e}^{-\mathcal{C}_1 t / \tau_\text{p}} )^{\mathcal{C}_2}.    
\end{align}  
\end{subequations}
The marginalization along $\xi$ gives the PDF of the particle velocity $U_\text{p}(t)$ in FHHS,
\begin{align}\label{eq:f_U-sol}
    f_{U}(u_\text{p};t) =\int \frac{1}{\eta} f_{\Xi }(\xi) \delta\left( \frac{u_\text{p}}{\eta}-\xi \right) \text d\xi = \frac{1}{\eta}f_{\Xi}\left( \frac{u_\text{p}}{\eta} \right).
\end{align}
In accordance with~\eqref{eq: FHHS_T_definition}, the variance of this PDF, $\sigma_U^2(t)$, is proportional to granular temperature $T(t)$. If $\Xi$ were Gaussian, $f_{\Xi}\sim \mathcal{N}(0,\sigma_{\Xi}^2)$, then 
\begin{align}\label{eq: FHHS_mean_up_and_sigma_up}
    \bar{U}_\text{p}(t) = {\bar{U}_\text{p}}(0) \text{e}^{-t / \tau_\text{p}} \quad\text{and}\quad
    \sigma_U^2(t)  = \sigma_{\Xi}^2 ( 1 - \text{e}^{-\mathcal{C}_1 t / \tau_\text{p}} )^{2\mathcal{C}_2}.
\end{align}

\begin{figure}[hbt!]
    \centering
    \includegraphics[width=0.5\textwidth]{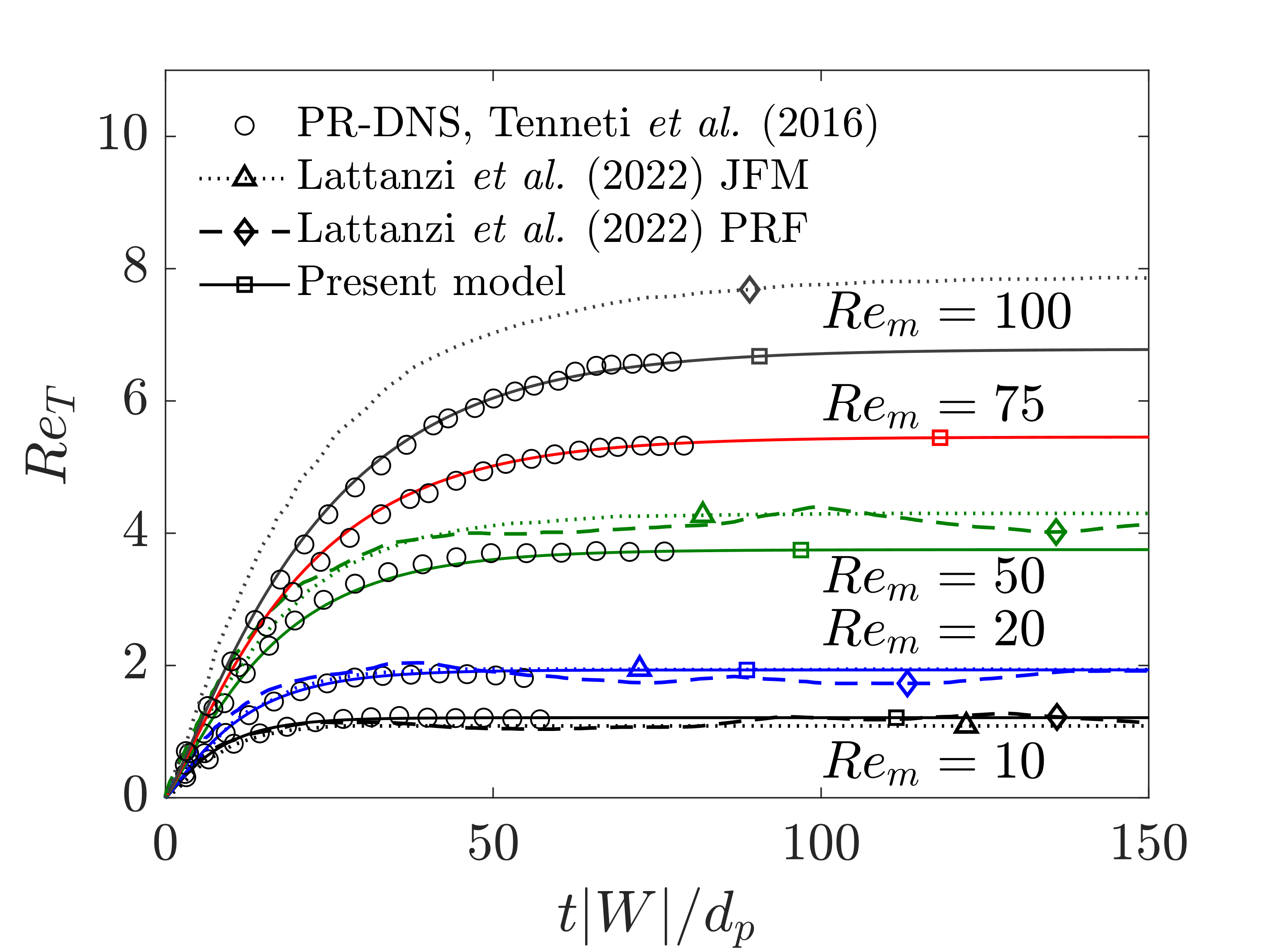}
    \caption[]{Temporal evolution of the thermal Reynolds number $Re_T(t)$ alternatively predicted with our model~\eqref{eq: FHHS_mean_up_and_sigma_up} (solid lines with square symbol), the model of Ref.\cite{lattanzi2022fluid} (dotted lines with triangle symbol) and the Langevin approach of Ref.\cite{lattanzi2022stochastic} (dashed lines with diamond symbol), for several values of the mean Reynolds number $\Rey_\text{m}$. The open circles indicate the PR-DNS data~\cite{tenneti2016stochastic,lattanzi2022stochastic}, to which $\sigma_\Xi = \sigma_\Xi(\Rey_\text{m})$ and $\mathcal C_1 = \mathcal C_1(\Rey_\text{m})$ are fitted (Fig.~\ref{fig: PR_DNS_FHCS_sigma_alpha}). The average volume fraction and density ratio are $\omega=0.1$ and $\rho_\text{p}/\rho_\text{f}=100$ respectively. For all cases $\mathcal{C}_2=1.2$ and $\tau_\text{p} = 0.14$. 
    }
    \label{fig: PR_DNS_FHCS_comparison}
\end{figure}

\begin{figure}[hbt!]
    \centering
    \includegraphics[width=0.4\textwidth]{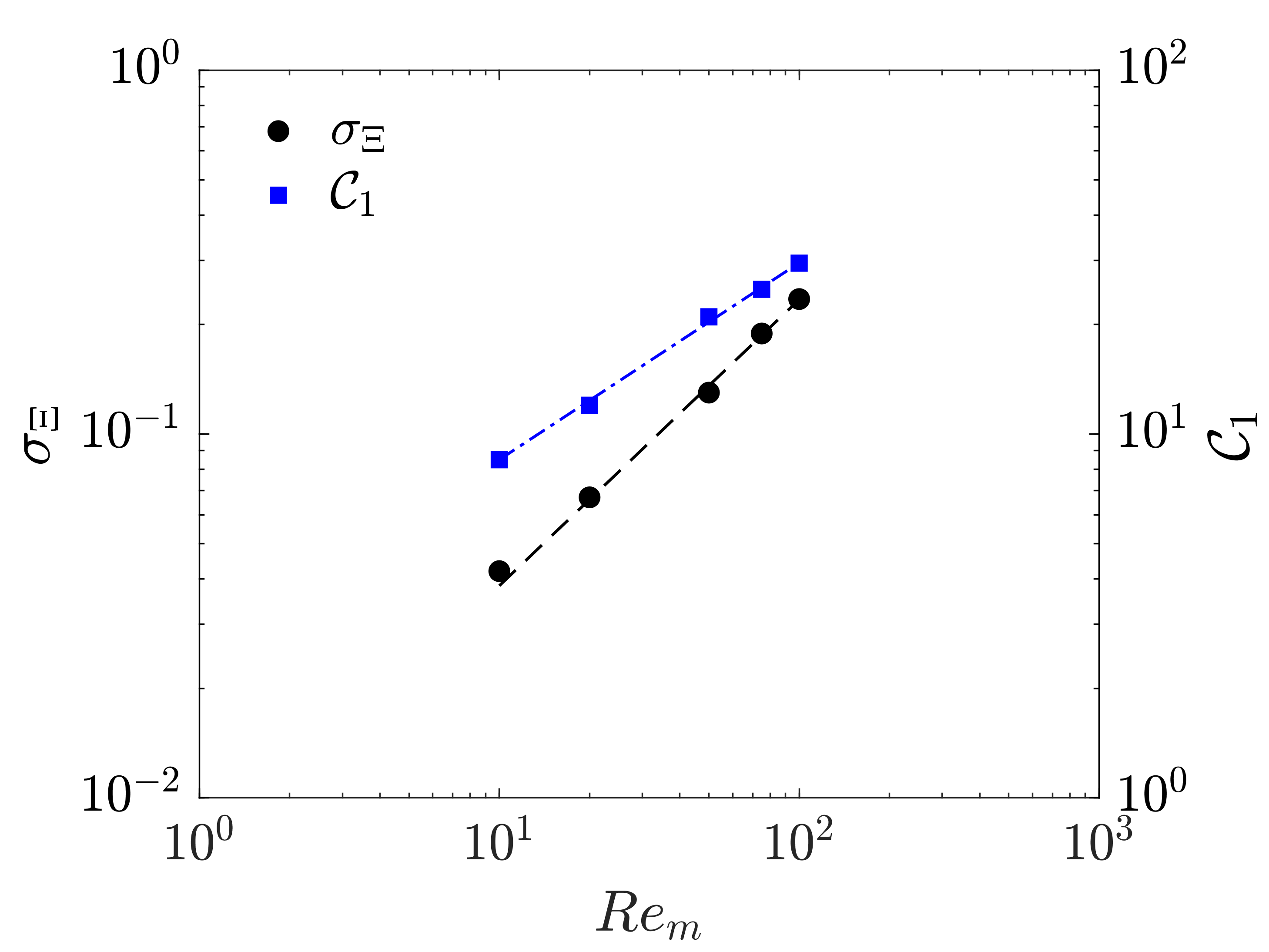}
    \caption[]{Functional dependencies of the model parameters $\sigma_{\Xi}$ and $\mathcal{C}_1$ on the mean Reynolds number $\Rey_\text{m}$. The squares and circles indicate the values of $\sigma_{\Xi}$ and $\mathcal{C}_1$ obtained via fitting~\eqref{eq: FHHS_mean_up_and_sigma_up} to the PR-DNS data \cite{tenneti2016stochastic} for $\rho_\text{p}/\rho_\text{f}=100$ and $\omega = 0.1$. These data are represented by $\log(\sigma_{\Xi}) = 0.06258 \log(\Rey_\text{m}) + \log(0.7866)$ and $\log(\mathcal{C}_1) = 2.446 \log(\Rey_\text{m}) + \log(0.5411)$, with the coefficient of determination $R^2=0.998$. }
    \label{fig: PR_DNS_FHCS_sigma_alpha}
\end{figure}

Figure~\ref{fig: PR_DNS_FHCS_comparison} exhibits the granular temperature predicted with this solution, $T(t) = \sigma_U^2 / 3$, for $\tau_\text{p} = 0.14$, $\rho_\text{p}/\rho_{\text{f}}=100$ and $\omega=0.1$~\cite{lattanzi2022stochastic}. 
The functional dependencies $\sigma_\Xi = \sigma_\Xi(\Rey_\text{m})$ and $\mathcal C_1 = \mathcal C_1(\Rey_\text{m})$ are obtained by fitting the predictions of $T(t)$ from~\eqref{eq: FHHS_mean_up_and_sigma_up} to the PR-DNS data from \cite{tenneti2016stochastic}.
This fitting reveals that both $\sigma_{\Xi}$ and $\mathcal{C}_1$ depend linearly in a log-log scale 
on $\Rey_\text{m}$ (Fig.~\ref{fig: PR_DNS_FHCS_sigma_alpha}), whereas $\mathcal{C}_2$ is constant ($\mathcal{C}_2=1.2$).
Our model matches the PR-DNS data better than its Langevin alternatives \cite{lattanzi2022stochastic,  lattanzi2022fluid}. 
It captures both the early growth and asymptotic values of $T(t)$, while the models of Refs.\cite{lattanzi2022fluid} and \cite{lattanzi2022stochastic} underestimate the steady values of $T(t)$ and predict the unphysical/unobserved long-term fluctuations in $T(t)$, respectively.

\subsection{Phenomenology of granular-temperature dynamics}

\begin{figure}[hbt!]
    \centering
    \includegraphics[width=0.4\textwidth]{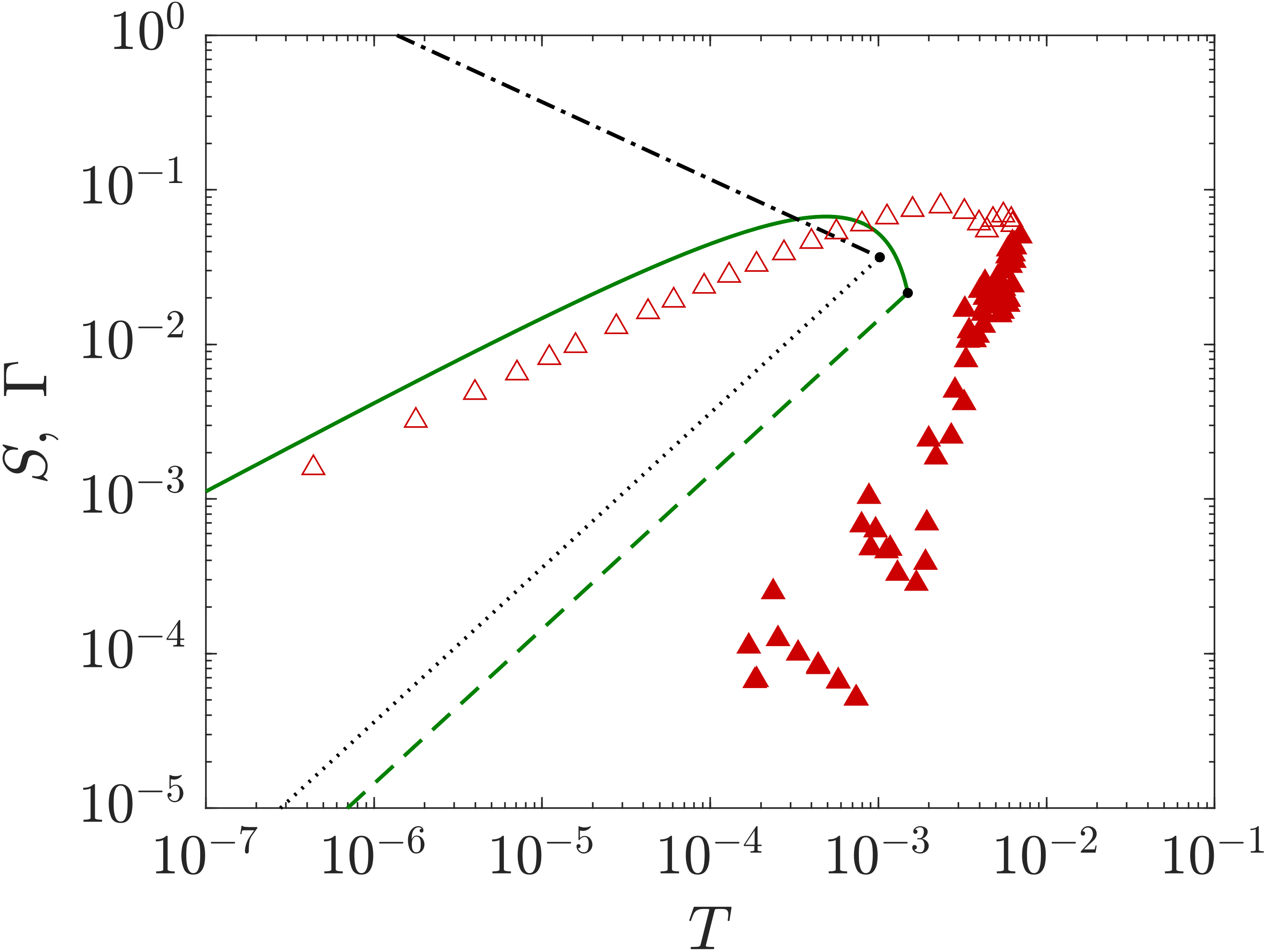}
    \caption[]{The source, $S(T)$, and sink, $\Gamma(T)$, terms in the granular-temperature model~\eqref{eq: FHHS_Teq}. Our model~\eqref{eq:source-sink} yields the solid and dashed green lines for $S(T)$ and $\Gamma(T)$, respectively. The model of Ref.\cite{koch1999particle} has $S(T)$ and $\Gamma(T)$ depicted by the dash-dotted and dotted black lines, respectively. The PR-DNS estimates of $S(T)$ and $\Gamma(T)$\cite{tenneti2016stochastic} are represented by the non-filled and filled red triangles, respectively. The parameter values are $\Rey_\text{m}=20$, $\rho_\text{p}/\rho_\text{f}=100$ and $\omega=0.1$. }
    \label{fig: PR_DNS_FHCS_phaseSpace}
\end{figure}

It has been argued that granular temperature in a FHHS, $T(t)$, exhibits the source-and-sink dynamics \cite{koch1999particle, tenneti2016stochastic, lattanzi2022fluid}, 
\begin{align}
    \frac{\text{d}T}{\text{d}t} = S(T) - \Gamma(T).
    \label{eq: FHHS_Teq}
\end{align}
In Ref.\cite{koch1999particle}, the source is taken to be $S \sim T^{-1/2}$, which is singular at $T=0$, i.e., at initial time $t=0$.  
In contrast, the ensemble averaging of our Liouville model~\eqref{eq: FHHS_dudt_1} gives rise to the source $S(T)$ and sink $\Gamma(T)$ (see Appendix~\ref{sec: moment_models} for detail),
\begin{align}
    S =  \frac{ 2 }{ \sqrt{3} } \sigma_{\Xi} \phi T^{1/2} \quad\text{and}\quad
    \Gamma = \frac{2}{\tau_\text{p}}T,
   \label{eq:source-sink}
\end{align}
that behave well at $T=0$.
Figure~\ref{fig: PR_DNS_FHCS_phaseSpace} exhibits the curves $S(T)$ and $\Gamma(T)$ given by~\eqref{eq:source-sink} and by the model in Ref.\cite{koch1999particle}, in comparison with PR-DNS data~\cite{tenneti2016stochastic}. 
While the mismatch between both models and the data is noticeable, our model exhibit the correct qualitative behavior, while the model in Ref.\cite{koch1999particle} does not. 

\subsection{Particle-velocity distributions}

A Langevin counterpart to our Liouville model~\eqref{eq: FHHS_dudt_1} is a stochastic ODE $\text{d}{U_\text{p}} = - (1 / \tau_\text{p}) {U_\text{p}} \text{d}t + \sqrt{2 D} \text{d}W$, which gives rise to
the Fokker-Planck equation for the particle velocity PDF, $f_U(u_\text{p};t)$, 
\begin{align}
    \frac{\partial f_{U}}{\partial t} - \frac{\partial }{\partial u_\text{p}} \left( \frac{1}{\tau_\text{p}}u_\text{p}  f_{U}  \right) = \frac{\partial}{\partial u_\text{p}} \left( D  \frac{\partial f_{U}}{\partial u_\text{p}} \right).
    \label{eq: FHHS_FP}
\end{align}
Many studies focus on the numerical determination of the diffusion coefficient $D(u_\text{p},t)$ using the standard deviation of particle acceleration and its integral time scale~\cite{lattanzi2022stochastic} or directly fitting a correlation from PR-DNS data~\cite{tenneti2016stochastic}.
Our Liouville approach provides an alternative means to accomplish this task analytically.
Substituting $f_U(u_\text{p};t)$ from~\eqref{eq:f_U-sol} into~\eqref{eq: FHHS_FP} yields a second-order ODE for $D(u_\text{p},\cdot)$, whose solution is 
%
\begin{align}
    D = \eta \phi \sigma_\Xi^2 + \frac{K}{u_\text{p}} \exp\left(\frac{u_\text{p}^2}{2 \eta^2 \sigma_{\Xi}^2}\right),
    \label{eq: FHHS_D}
\end{align}
with $K$ an integration constant whose value is not constrain by the compatibility condition (see~\eqref{eq: diffusion_tensor_back_fxp} and~\eqref{eq: diffusion_tensor_back}), and it is an extra degree of freedom of the model. 
Here we choose $K=0$, such that the diffusion coefficient is simply $D = \eta \phi \sigma_\Xi^2$. 
Using the analytical relations derived previously, the diffusion tensor can be expressed in terms of the granular temperature as 
\begin{align} \label{eq: D_of_T}
    D = \frac{3}{\tau_\text{p}} \left[ \left( 1- \mathcal{C}_1\mathcal{C}_2 \right)T  +\frac{\mathcal{C}_1\mathcal{C}_2 \sigma_\Xi^{1/\mathcal{C}_2} }{3^{1/(2\mathcal{C}_2)}} T^{1-1/(2\mathcal{C}_2)}    \right].
\end{align}

\begin{figure}[hbt!]
    \centering
    \includegraphics[width=0.4\textwidth]{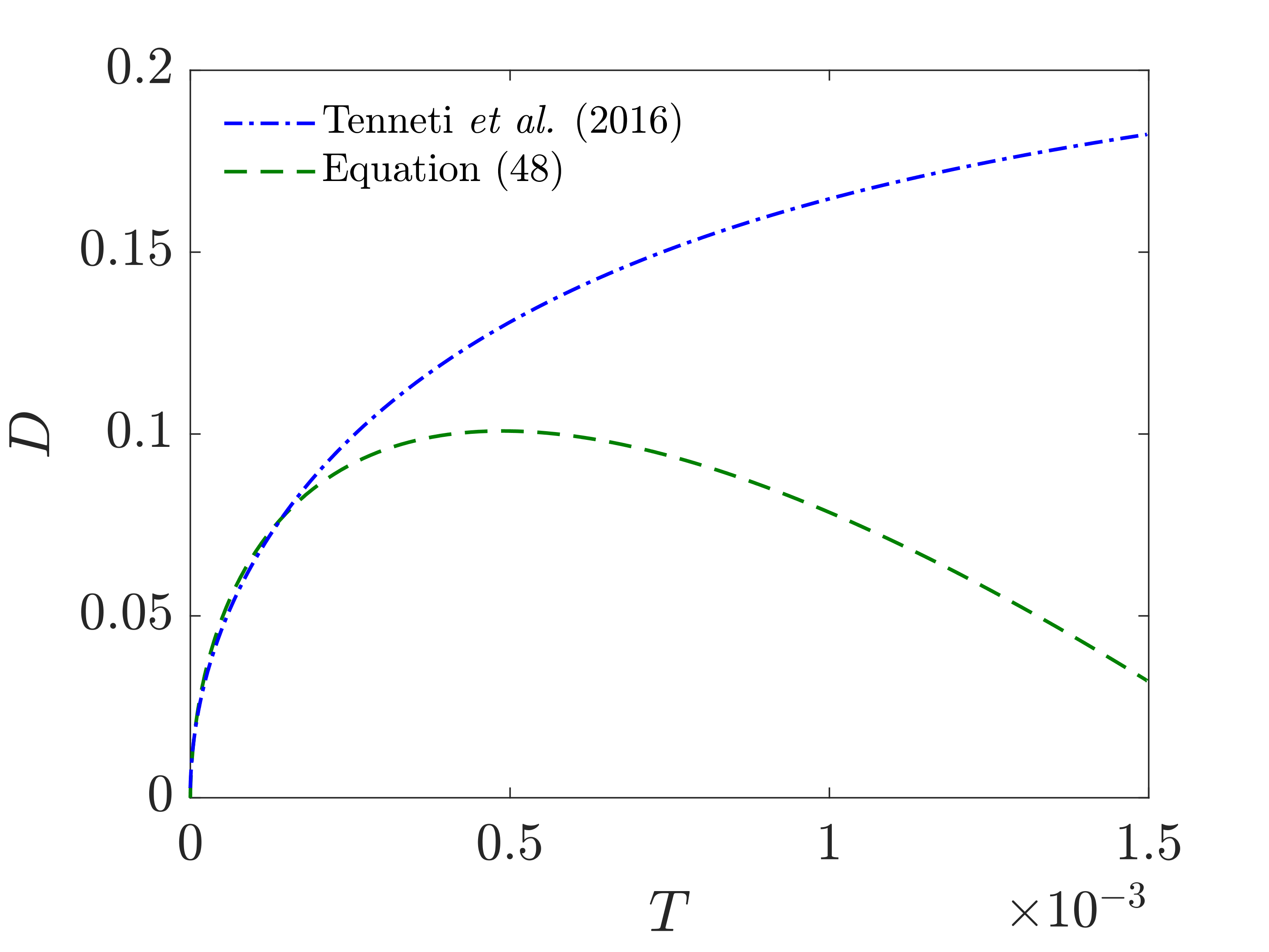}
    \caption[]{Diffusion coefficient in terms of the granular temperature in the model presented in Ref.\cite{tenneti2016stochastic} and the one developed here, equation~\eqref{eq: D_of_T}, for $\Rey_m=20$, $\omega=0.1$, $\rho_\text{p}/\rho_\text{f}=100$ and $\tau_\text{p}=0.14$.   }
    \label{fig: PR_DNS_FHCS_DiffComparison_R20r100p01}
\end{figure}

Models in the literature~\cite{tenneti2016stochastic,lattanzi2022stochastic} and, in general, numerical approximations of the diffusion coefficient in the Langevin approach are not consistently linked to Fokker-Planck equations and an equivalent PDF formulation is seldom investigated. 
Moreover, the full statistical description is provided by a PDF description~\eqref{eq:FHHS_jointsol}.
The model presented here with the Liouville approach leads to equivalent Langevin and Fokker-Planck formulations.
In Fig~\ref{fig: PR_DNS_FHCS_DiffComparison_R20r100p01} we compare the derived diffusion coefficient~\eqref{eq: D_of_T} with the model proposed in Ref.\cite{tenneti2016stochastic}, showing agreement at lower values of the granular temperature, linked to the evolution at the early times of the FHHS.
The model in Ref.\cite{tenneti2016stochastic} is based on a correlation that contains $8$ constants fitted from PR-DNS data.


\subsection{Miscellaneous musings} \label{sec:Miscellaneous_musings}

The advantage of the Liouville approach is obvious, the source and sink terms for the evolution of the granular temperature can be proposed starting from defining the standard deviations of the particle velocity~\eqref{eq: FHHS_mean_up_and_sigma_up} based on phenomenology.
Then, we proceed to find the source term $\phi$ going back to the characteristics that can be analytically described.
The model also leads a diffusion coefficient that can be computed accordingly by comparing the Fokker-Planck equation with the marginalized Liouville equation. 
Trying to formulate the problem from the Langevin perspective without previous knowledge of the diffusion coefficient and arrive at the expression in~\eqref{eq: FHHS_D} is a hard task.
The procedure can be used to develop a model based on the time evolution of $T$ reported in PR-DNS results by adjusting the parameters $\sigma_{\Xi}$, $\mathcal{C}_1$ and $\mathcal{C}_2$.

The temporal evolution of the granular temperature in FHHS is characterized by a source and sink behavior~\cite{tenneti2016stochastic} stabilizing at a constant value after an initial growth. To this purpose, the parameter $\sigma_{\Xi}$ is responsible for the steady value, and the coefficients $\mathcal{C}_1$ and $\mathcal{C}_2$ characterize the growth of the granular temperature at two distinguished time scales, at early and later times respectively. 
The Langevin models in Ref.\cite{lattanzi2020stochastic} correspond to ODE systems with a random coefficient in the Liouville approach with source terms not defined at the initial time, making the solution singular at $t=0$, corresponding to an initial infinite impulse of the particles (events moving in phase space). 
See the previous Sections~\ref{sec: PDF_solutions_position} and~\ref{sec: PDF_solutions_velocity} that reproduce results reported in Ref.\cite{lattanzi2020stochastic}.
Another model where the solution is not defined at the initial time is the one presented in Ref.\cite{koch1999particle}, where the granular temperature grows infinitely at the initial time.
In our model, the initial impulse in the granular temperature is bounded and governed by the parameter $\mathcal{C}_2$, leading to an ODE system whose solution is bounded at $t=0$.

\section{Conclusions} \label{sec: conclusions}

A framework to formulate stochastic descriptions of particle-laden flows based on a Liouville PDF governing equation introduced.
A connection between Langevin dynamical systems and the new Liouville approach allows to analytically treat the solution with the method of characteristics and circumvent numerical difficulties related to It\^o calculus and the numerical resolution of the Fokker-Planck equation. The Liouville approach is based on expressing stochastic processes with random coefficients constants in time and additional source terms. The problem is then governed by the Liouville equation that can be derived with the method of distributions in an augmented dimensionality that includes the added random coefficients, the particle position and particle velocity. With proper selection of the random coefficients and source terms, both Langevin and Liouville approaches coincide. This implies that the Liouville equation marginalized along the extra dimension of the random coefficients leads to the Fokker-Planck equation of an equivalent Langevin system.

The Liouville approach is compared with canonical Eulerian-Lagrangian models of particle-laden flows formulated with Langevin equations used to obtain the time evolution of particle dispersion and granular temperature in the literature. We develop a full description of the statistics with the Liouville approach, that analytically describes exact results of the PDFs of the particle position and velocity as well as their moments.

The potential of the Liouville approach to develop modeling of particle statistics is assessed by proposing a phenomenological model to FHHS. Based on the analytical character of the Liouville approach, the sink and source terms that govern the time evolution of the granular temperature can be represented in the Liouville approach by selecting a source term that depends on the non-dimensional parameters of the problem. This leads to matching of PR-DNS results more accurately than previous proposed numerical and analytical models based on the Langevin approach. The FHHS model presented here also eliminates the singularity at the initial time in the granular temperature equation, that leads to an ill-posed problem that contains an initial infinite impulse. Previous models in the literature did not offer descriptions of the granular temperature without such singularity. This singularity at the initial time can be interpreted as an initial infinite impulse by virtue of the Liouville approach that links events with possible particles. Such events are associated with deterministic characteristic equations that are derived from the Liouville equation with the method of characteristics. The hyperbolicity of the Liouville equation is therefore advantageous as opposed to the advection-diffusion character of the Fokker-Planck equation. The model presented here is well posed at the initial time and describes the early evolution of the granular temperature accurately as compared with PR-DNS. Additionally, the Liouville approach may be used to find the diffusion coefficient and with it other descriptions of the same problem based on the Fokker-Planck or Langevin equations.

The Liouville approach may be used to describe other Langevin dynamical systems in general, making its numerical treatment simpler. It is not bounded by Gaussianity and does not rely on sampling nor requires closure. It opens new possibilities in proposing reduced models that capture statistics obtained with PR-DNS. Moreover, because events and their probability to occur are traced deterministically, each event has a deterministic equation given by the characteristics that can be physically interpreted as a possible particle in the system. Then each event is associated with a physical particle and how likely to occur is that particle in the system. Therefore it provides a connection between the statistics and the particle physics.

Additionally, moment equations provide a reduced description, which may lead to closure challenges depending on the non-linearity of the particle equations. In the results presented here, the moment equations are found in closed form.

Extensions of this work include accounting for mass and energy exchange between the two phases, as well as the development of stochastic models for particle statistics in flows with particle-shock interactions~\cite{sen2018evaluation,das2018strategies,das2018metamodels}.


\section*{Funding}

This work was supported by the Air Force Office of Scientific Research under award numbers FA9550-19-1-0387 and FA9550-21-1-0381, and by San Diego State University Graduate Fellowship. 

\section*{Declaration of interests} 

The authors report no conflict of interest.



\section*{Data Availability Statement}

The data that support the findings of this study are available upon reasonable request.

\appendix

\section{Moment Models} 
\label{sec: moment_models}


With the Liouville approach, descriptions of moments of the particle phase variables may also be derived using the systems~\eqref{eq: position_ODE_liouville_general} or~\eqref{eq: velocity_ODE_liouville_general}. These include the variance of particle positions (particle dispersion), the variance of particle velocity (granular temperature) and higher moments as the third and fourth moment of the particle position less reported in the literature. To write a general moment model based on the systems of ODEs with a random coefficient described previously, we first rewrite the ODE systems in terms of the vector of state variables $\boldsymbol{q}$. Then, both systems~\eqref{eq: position_ODE_liouville_general} and~\eqref{eq: velocity_ODE_liouville_general} can be written as
\begin{align}
    \frac{\text{d}\boldsymbol{q}}{\text{d}t} = \boldsymbol{g}(\boldsymbol{q}) ,
    \label{eq: general_ODEsystem_Xi}
\end{align}
where $\boldsymbol{q}=(\boldsymbol{x}_\text{p}, \ \Xi) \in \mathbb{R}^4$ for the model in~\eqref{eq: position_ODE_liouville_general} and $\boldsymbol{q}=(\boldsymbol{x}_\text{p}, \ \boldsymbol{u}_\text{p}, \ \Xi) \in \mathbb{R}^7$ for~\eqref{eq: velocity_ODE_liouville_general} with the corresponding function with right hand side terms here taken as $\boldsymbol{g}$, and considering a single random coefficient $\Xi$.
Then, we employ a Reynolds decomposition of the random variables $(\cdot)=\overline{(\cdot)}+(\cdot)^\prime$. Substituting in~\eqref{eq: general_ODEsystem_Xi} and averaging, we arrive to the first two moment equations
\begin{subequations} \label{eq: moment_models_q}
\begin{align}
    \frac{\text{d} \overline{q}_i}{\text{d}t} &= \overline{g}_i , 
    \label{eq: moment_models_q_mean} \\
    \frac{\text{d}{\overline{{q_i^\prime}{q_j^\prime}}}}{\text{d}t} &= \overline{q_j^\prime g_i^\prime} + \overline{q_i^\prime g_j^\prime}, 
    \label{eq: moment_models_q_cm2} 
\end{align}
\end{subequations}
where unclosed terms appear as the average of the non-linear function $\boldsymbol{g}$ is in principle unknown, as well as second order moments involving it. Closures for such terms may be based on \textit{a priori} computations~\cite{davis2017sparse} or expansions~\cite{dominguez2023closed,dominguez2023sparser}. If the right hand side of the ODE system is composed by a product of non-linear functions of the state variables, $\boldsymbol{q}$, statistical moments of order higher appear in each moment equation, defining a classic closure problem. Take for exmaple $\boldsymbol{g}(\boldsymbol{q})=\boldsymbol{r}(\boldsymbol{q}) \circ \boldsymbol{h}(\boldsymbol{q})$, then the system~\eqref{eq: moment_models_q} is rewritten as
\begin{subequations} \label{eq: moment_models_qProc}
\begin{align}
    \frac{\text{d} \overline{q}_i}{\text{d}t} &= \overline{r}_i \overline{h}_i +\overline{r_i^\prime h_i^\prime}, 
    \label{eq: moment_models_qProc_mean} \\
    \frac{\text{d}{\overline{{q_i^\prime}{q_j^\prime}}}}{\text{d}t} &= \overline{q_j^\prime h_i^\prime}\overline{r}_i + \overline{q_j^\prime r_i^\prime} \ \overline{h}_i + \overline{q_i^\prime h_j^\prime}\overline{r}_j + \overline{q_i^\prime r_j^\prime} \ \overline{h}_j + \overline{q_j^\prime h_i^\prime r_i^\prime} + \overline{q_i^\prime h_j^\prime r_j^\prime},
    \label{eq: moment_models_qProc_cm2} 
\end{align}
\end{subequations}
with $\boldsymbol{r}$ and $\boldsymbol{h}$ two non-linear functions. This is for example is the case if a correction of the Stokes drag $\mathcal{F}$ based on the relative velocity is used in~\eqref{eq: velocity_ODE_liouville_general}. By combining closures based on expansions with truncation of higher order terms and a splitting algorithm, in Ref.\cite{dominguez2023sparser}, the first two moments were computed in closed form.

When applied to~\eqref{eq: FHHS_dudt_1}, the method of moments yields 
\begin{subequations} \label{eq: FHHS_T_eqs}
\begin{align}
    \frac{\text{d}T}{\text{d}t} &= -\frac{2}{\tau_\text{p}}T +\frac{2}{3} \overline{\Xi^\prime U_\text{p}^\prime} \phi, 
    \label{eq: FHHS_T_eqs_T} \\
    \frac{\text{d}\overline{\Xi^\prime U_\text{p}^\prime}}{\text{d}t} &= -\frac{1}{\tau_\text{p}}\overline{\Xi^\prime U_\text{p}^\prime} + \sigma_{\Xi}^2 \phi.
    \label{eq: FHHS_T_eqs_aup}
\end{align}  
\end{subequations}
The solution of the second ODE is
\begin{subequations} \label{eq: FHHS_T_aup_result}
\begin{align}
    \overline{\Xi^\prime U_\text{p}^\prime} = \sigma_{\Xi}^2 ( 1 - \text{e}^{-\mathcal{C}_1 t / \tau_\text{p}} )^{\mathcal{C}_2}.
    \label{eq: FHHS_aup_result}
\end{align}  
\end{subequations}
Since $T = \sigma_U^2 / 3$, with $\sigma_U^2$ given by~\eqref{eq: FHHS_mean_up_and_sigma_up}, substituting this result into~\eqref{eq: FHHS_T_eqs_T}, yields the source $S$  and sink $\Gamma$ terms in~\eqref{eq:source-sink}.

\section{Analytical solutions based on the moment model}
\label{sec: moment_solutions}

\subsection{Particle trajectory}
\label{sec: position_moments}

The moment equations of the one-dimensional version of~\eqref{eq: position_ODE_liouville_general} with a single random variable $Z = \Xi$ read as
\begin{subequations} \label{eq: position_moments_sol}
\begin{align}
    \frac{\text{d}\overline{X}_\text{p}}{\text{d}t} &= \ u_\text{p} + \overline{\Xi} \varphi, 
    \label{eq: position_moments_mean_x_sol} \\
    \frac{\text{d}\overline{{X_\text{p}^\prime}^2}}{\text{d}t} &= 2  \overline{X_\text{p}^\prime U_\text{p}^\prime} + 2\overline{ \Xi^\prime X_\text{p}^\prime } \varphi, 
    \label{eq: position_moments_cm2_x_sol} \\
    \frac{\text{d}\overline{\Xi^\prime X_\text{p}^\prime }}{\text{d}t} &= \overline{{\Xi^\prime}^2}\varphi.
    \label{eq: position_moments_xa_sol}
\end{align}
\end{subequations}
Considering constant and deterministic velocity $u_\text{p}$, one has $u_\text{p}=\overline{u}_\text{p} $ and $u_\text{p}^\prime=0$. The impulse initial condition, given by a Dirac delta function located at zero, gives $\overline{X}_{\text{p}_0}=\overline{{X_\text{p}^\prime}_0^2}=0$. Also, initially the random coefficient and the particle locations are statistically independent such that $\overline{\Xi^\prime X_\text{p}^\prime}_0=\sigma_\Xi \sigma_{{X_\text{p}}_0}=0$. With this initial condition, the system~\eqref{eq: position_moments_sol} leads to the following solution
\begin{subequations} \label{eq: position_moments_solution_general}
\begin{align}
    \overline{X}_\text{p} &= u_\text{p} t+\overline{\Xi}\sqrt{2 D t}, 
    \label{eq: position_moments_solution_general_mean_x} \\
    \overline{{{X_\text{p}^\prime}^2}} &=  2 \overline{{\Xi^\prime}^2} D t , 
    \label{eq: position_moments_solution_general_cm2_x} \\
    \overline{\Xi^\prime X_\text{p}^\prime } &=  \overline{{\Xi^\prime}^2} \sqrt{2 D t},
    \label{eq: position_moments_solution_general_xa}
\end{align}
\end{subequations}
where if $\overline{\Xi}=0$ and $\overline{{\Xi^\prime}^2}=1$, in concordance with the classic heat kernel solution where $\Xi \sim \mathcal{N}(0,1)$, the solution~\eqref{eq: position_moments_solution_general} leads to the classic result for the particle dispersion $\overline{{X_\text{p}^\prime}^2} = 2 D t$. The average position grows linear in time $\overline{X}_\text{p} =  u_\text{p} t$, and the correlation of position with the random coefficient is $\overline{\Xi^\prime X_\text{p}^\prime}= \sigma_{X_\text{p}}$. The characteristic in equation~\eqref{eq: position_characteristics_1D_solution_X} can be recasted in the form $\hat{x}_\text{p} = \overline{X}_\text{p} + \xi \sigma_{X_\text{p}} $ according to~\eqref{eq: position_moments_solution_general}. This linear relation indicates that for each value of $\xi$, a quantity $\xi \sigma_{X_\text{p}}$ is added to the average path and causes that corresponding trajectory to deviate from the average.

The general expression for the average particle position can be also obtained by averaging of the characteristic equation $\hat{x}_\text{p}(\xi,t)$ with respect to $\xi$, which corresponds to the application of the law of the unconscious statistician (LOTUS)~\cite{flury2013first}. Making use of~\eqref{eq: position_characteristics_1D_solution_X}, with the impulse located at zero $\hat{x}_\text{p}^0=0$, one has
\begin{align}
\begin{split}
    \overline{X}_\text{p} &= \int \hat{x}_\text{p}(\xi,t) f_\Xi(\xi)d\xi = \int \left( \hat{x}_\text{p}^0 + u_\text{p} t +\xi \sqrt{2 D t} \right)f_\Xi(\xi)d\xi  
    =  u_\text{p} t + \overline{\Xi}\sqrt{2 D t} 
    \end{split}
\end{align}
and for the $n$-th central moment 
\begin{align}
    \overline{{X_\text{p}^\prime}^n} &= \int (\hat{x}_\text{p}(\xi,t) - \overline{X}_\text{p})^n f_\Xi(\xi)d\xi = \overline{{\Xi^\prime}^n} \left( 2 D t \right)^{n/2} ,
    \label{eq: position_cm3}
\end{align}
for $n\geq 2$. Notice that the even moments are non-zero for any of the chosen distributions for $\Xi$ ($\mathcal{N}$, $\mathcal{U}$ or $\mathcal{T}$) whereas the odds moments are zero for symmetric distributions. These result generalize those reported in Ref.\cite{lattanzi2020stochastic} for non-Gaussian white noise.

\subsection{Particle velocity}
\label{sec: velocity_moments}

\begin{figure}[h!]
    \centering
    \subfloat[]{
    \label{fig: VL_xpxp_upup}
    \includegraphics[width=0.4\textwidth]{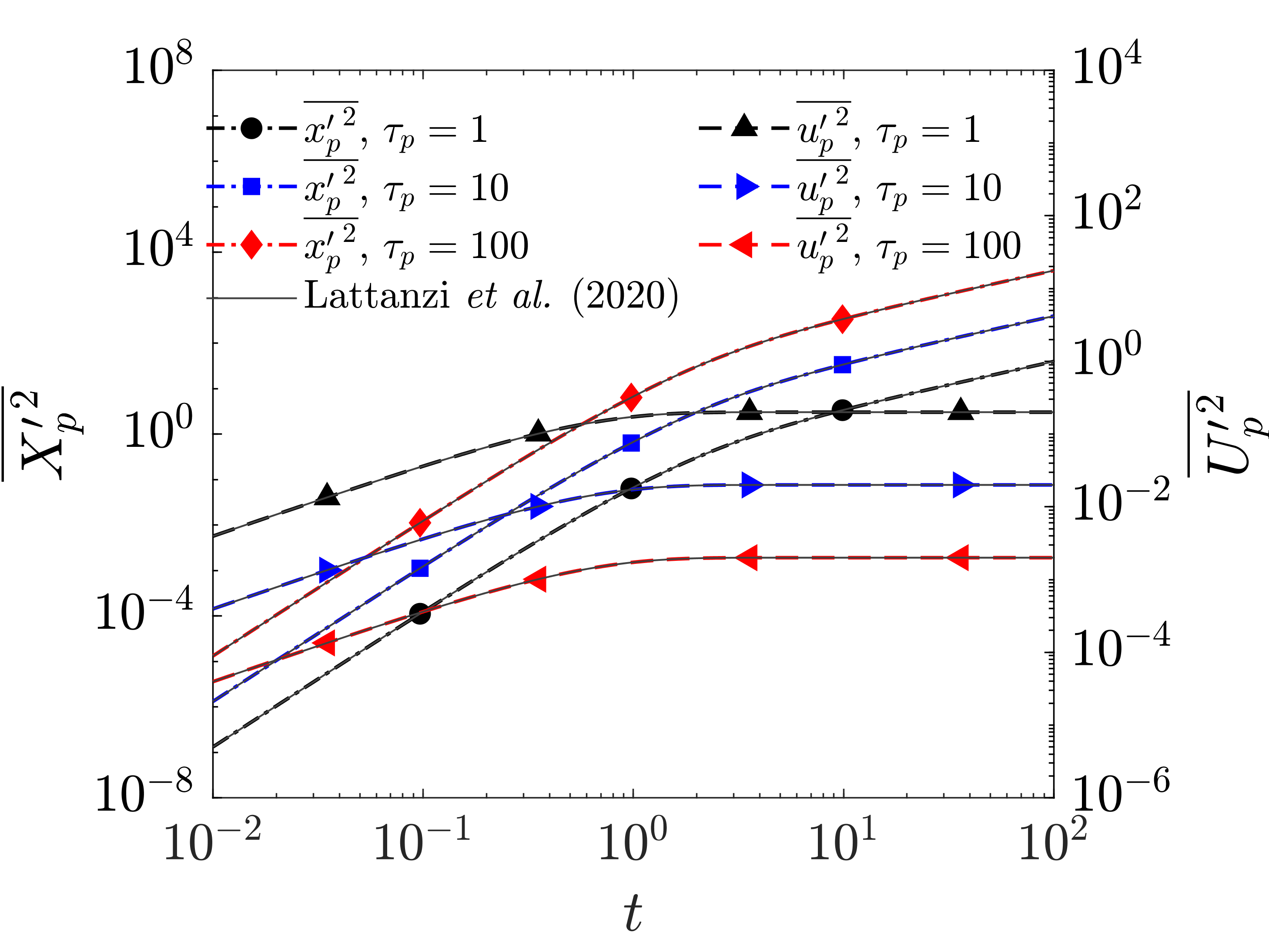}}
    \hfill	
    \subfloat[]{
    \label{fig: VL_xpup_xpa_upa}
    \includegraphics[width=0.4\textwidth]{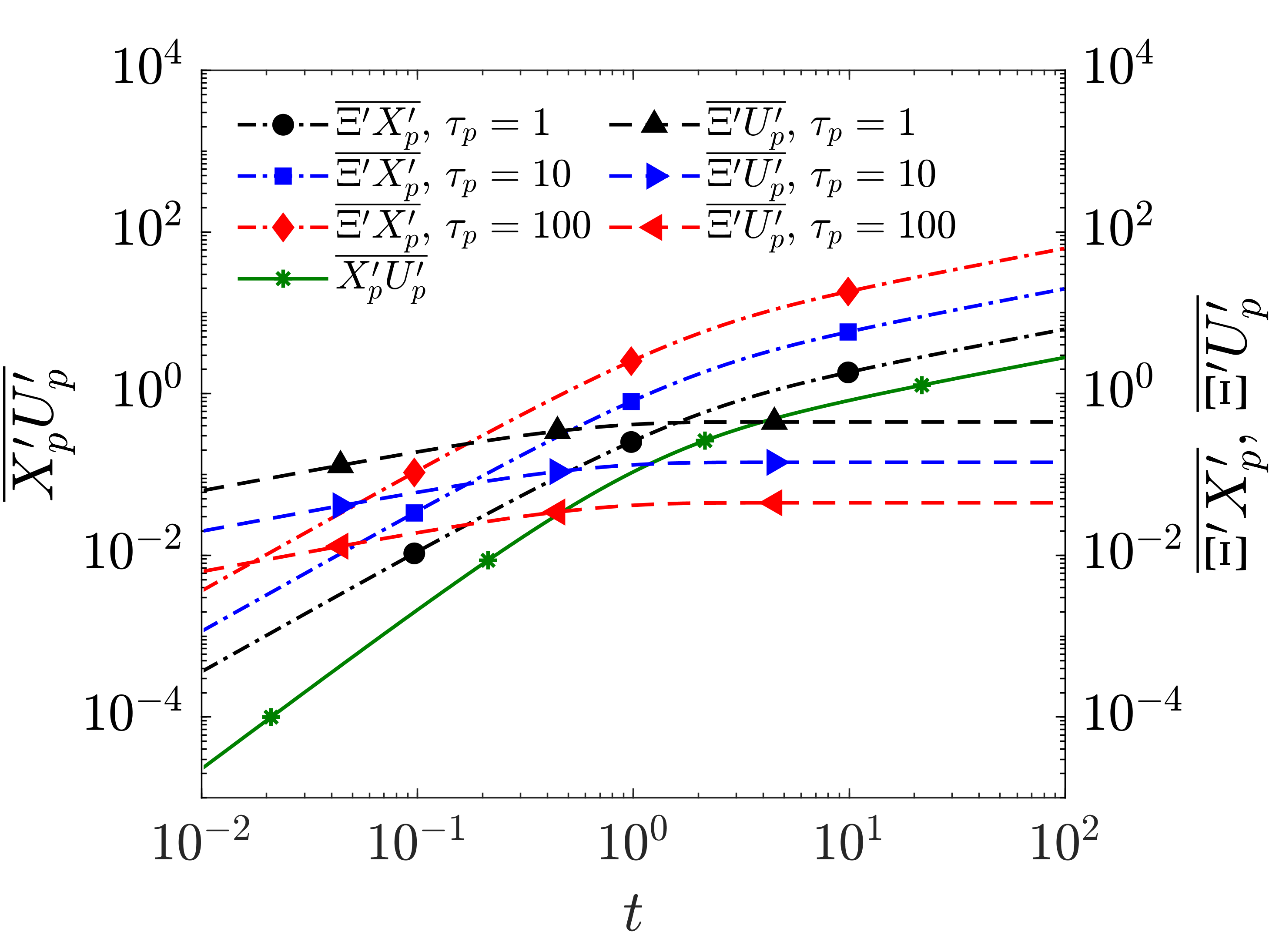}} 
    \caption[]{Moments of the particle position and velocities according to the solutions~\eqref{eq: velocity_SOLmoments}. In (a) solutions of the particle position and velocity variance which coincides with solutions reported in Ref.\cite{lattanzi2020stochastic} and (b) additional moments correlating particle positions with velocities and the random coefficient with particle variables. The cases correspond to particle time constants $\tau_\text{p}=[1, \ 10, \ 100]$ and $(\tau_p D)^{-1}=5$.}
    \label{fig: velocity_moments}
\end{figure}

The moment equations of the system~\eqref{eq: velocity_ODE_liouville_sol} read as
\begin{subequations} \label{eq: velocity_moments}
\begin{align}
    \frac{\text{d}\overline{X}_\text{p} }{\text{d}t} &= \tau_\text{p} \overline{U}_\text{p} + \overline{\Xi}\varphi, 
    \label{eq: velocity_moments_mean_xp} \\
    \frac{\text{d}\overline{U}_\text{p} }{\text{d}t} &= -\overline{U}_\text{p} + \overline{\Xi}\phi, 
    \label{eq: velocity_moments_mean_up} \\
    \frac{\text{d}\overline{{X^\prime_\text{p}}^2}}{\text{d}t} &= 2 \tau_\text{p} \overline{X_\text{p}^\prime U_\text{p}^\prime} + 2 \varphi \overline{\Xi^\prime X_\text{p}^\prime},
    \label{eq: velocity_moments_cm2_xp} \\
    \frac{\text{d}\overline{X^\prime_\text{p} U_\text{p}^\prime}}{\text{d}t} &=  \tau_\text{p} \overline{{U_\text{p}^\prime}^2} - \overline{X_\text{p}^\prime U_\text{p}^\prime} +  \varphi \overline{\Xi^\prime U_\text{p}^\prime } + \phi \overline{\Xi^\prime X_\text{p}^\prime},
    \label{eq: velocity_moments_xpup} \\
    \frac{\text{d}\overline{{U_\text{p}^\prime}^2}}{\text{d}t} &=  -2\overline{{U_\text{p}^\prime}^2} + 2\phi \overline{\Xi^\prime U_\text{p}^\prime } ,
    \label{eq: velocity_moments_cm2_up} \\
    \frac{\text{d}\overline{\Xi^\prime X_\text{p}^\prime }}{\text{d}t} &=  \tau_\text{p}\overline{\Xi^\prime U_\text{p}^\prime } + \overline{{\Xi^\prime}^2} \varphi,
    \label{eq: velocity_moments_xpa} \\
    \frac{\text{d}\overline{\Xi^\prime U_\text{p}^\prime}}{\text{d}t} &= -\overline{\Xi^\prime U_\text{p}^\prime}+  \overline{{\Xi^\prime}^2} \phi,
    \label{eq: velocity_moments_upa} 
    \end{align}
\end{subequations}
where the equations are closed because in this reference frame the flow velocity does not appear in the average velocity equation and the Stokes drag has been assumed, $\mathcal{F}=1$. The solution to the system~\eqref{eq: velocity_moments} considering $\overline{\Xi}=0$ and $\sigma_{\Xi}=1$, according to $\Xi \sim \mathcal{N}(0,1)$, and all other first and second moments zero at the initial time except ${\overline{U}_\text{p}}_0=v_0$, as well as statistical independence of the variables initially, is given by
\begin{subequations} \label{eq: velocity_SOLmoments}
\begin{align}
    \overline{X}_\text{p}  &= \tau_\text{p} v_0\left( 1-e^{-t} \right), 
    \label{eq: velocity_SOLmoments_mean_xp} \\
    \overline{U}_\text{p} &=  v_0 e^{-t} 
    \label{eq: velocity_SOLmoments_mean_up} \\
    \overline{{X^\prime_\text{p}}^2} &= \tau_\text{p}^2 D\left( 2t-3+4e^{-t}-e^{-2t} \right),
    \label{eq: velocity_SOLmoments_cm2_xp} \\
    \overline{X^\prime_\text{p} U_\text{p}^\prime}&=  \tau_\text{p} D \sqrt{\left(1-e^{-2t} \right) \left(2t-3+4e^{-t}-e^{-2t}  \right)},
    \label{eq: velocity_SOLmoments_xpup} \\
    \overline{{U_\text{p}^\prime}^2} &=  D \left( 1-e^{-2t} \right) ,
    \label{eq: velocity_SOLmoments_cm2_up} \\
    \overline{\Xi^\prime X_\text{p}^\prime} &= \sqrt{ \tau_\text{p}^2 D \left( 2t-3+4e^{-t}-e^{-2t} \right)  }  ,
    \label{eq: velocity_SOLmoments_xpa} \\
    \overline{\Xi^\prime U_\text{p}^\prime}&= \sqrt{D \left( 1-e^{-2t} \right)}.
    \label{eq: velocity_SOLmoments_upa} 
\end{align}
\end{subequations}
The averages particle position and velocity as well as the variances of particle positions $\overline{{X_\text{p}^\prime}^2}$ and velocities $\overline{{U_\text{p}^\prime}^2}$ is exact to the results reported in Ref.\cite{lattanzi2020stochastic} for the velocity Langevin model (see Figure~\ref{fig: VL_xpxp_upup}). In addition, the solution~\eqref{eq: velocity_SOLmoments} includes the second moments of variables combined such as the correlation of particle positions and velocities, $\overline{X_\text{p}^\prime U_\text{p}^\prime}$, which have not been previously reported in the literature for this particular model. For completeness, we show in Figure~\ref{fig: VL_xpup_xpa_upa} the rest of the second moments of the particle phase which are related to the standard deviation of the particle position and velocity by $\overline{X_\text{p}^\prime U_\text{p}^\prime}=\sigma_X \sigma_U$, $\overline{\Xi^\prime X_\text{p}^\prime }=\sigma_{\Xi}\sigma_{X}$ and $\overline{\Xi^\prime U_\text{p}^\prime}=\sigma_{\Xi}\sigma_{U} $.

The general system~\eqref{eq: velocity_moments}, may be used for any other initial conditions or correlations of the fluctuations by selecting $\overline{\Xi}$ and $\overline{{\Xi^\prime}^2}$ differently. 
For example, if the particle velocity is initialized with a Maxwellian distribution $\mathcal{N}(0,D)$, such that $\overline{{U_\text{p}^\prime}^2_0}=D$, the system~\eqref{eq: velocity_moments} reproduces the classic ballistic–diffusive result of Taylor~\cite{taylor1922diffusion} where $\overline{{X_\text{p}^\prime}^2}=2\tau_\text{p}^2 D\left( t-1+e^{-t} \right)$. This is also for $\overline{\Xi}=0$ and $\overline{{\Xi^\prime}^2}=1$.

Higher moments are also found with this moment model. Using the characteristics equations~\eqref{eq: velocity_characteristics_solution_Up} and applying the LOTUS, one may find the following general expressions for the $n$-th central moments
\begin{subequations}
\begin{align}
    \overline{{X_\text{p}^\prime}^n} &= \overline{{\Xi^\prime}^n} \left[ \tau_\text{p}^2 D\left( 2t-3+4e^{-t}+e^{-2t} \right) \right]^{n/2}, \\
    \overline{{U_\text{p}^\prime}^n} &= \overline{{\Xi^\prime}^n} \left[ D \left(1-e^{-2t} \right) \right]^{n/2}. 
\end{align}
\end{subequations}

\section{Derivation of analytical solutions}
\label{app:derivations}

The method of characteristics, i.e., the mapping of the fixed (Eulerian) coordinate system in the phase space, $x_\text{p}$, onto the moving (Lagrangian) coordinate system in the phase space, $\hat x_\text{p}(t)$, transforms the PDE~\eqref{eq:3.3} into a system of ODEs
\begin{subequations}    \label{eq: position_characteristics_1D}
\begin{align}
    \frac{\text{d}\hat{x}_\text{p}}{\text{d}t} = &\; u_\text{p} + \xi \sqrt{\frac{D}{2 t}}, \qquad \hat x_\text{p}(0) = \hat{x}_\text{p}^0; \\ 
    \frac{\text{d} f_{X \Xi}}{\text{d}t} = & \; 0, \qquad f_{X \Xi}(\hat{x}_\text{p}(0),\xi;0)=f_{\Xi}(\xi)\delta(\hat{x}_\text{p}^0).
\end{align}
\end{subequations}
The solution to the characteristics~\eqref{eq: position_characteristics_1D} is
\begin{subequations} \label{eq: position_characteristics_1D_solution}
\begin{align}
    \hat{x}_\text{p} &= \hat{x}_\text{p}^0 + u_\text{p} t + \xi \sqrt{2 D t }, 
    \label{eq: position_characteristics_1D_solution_X} \\
    f_{X \Xi} &= f_{X \Xi}^0.
\label{eq: position_characteristics_1D_solution_fax}
\end{align}
\end{subequations}
After substituting $\hat{x}_\text{p}^0$ from~\eqref{eq: position_characteristics_1D_solution_X} we arrive to~\eqref{eq: position_joint_aXp_1}.

The characteristics of the Liouville equation~\eqref{eq: velocity_Liouville_3D} are given by
\begin{subequations} \label{eq: velocity_characteristics}
\begin{align}
    \frac{\text{d}\hat{x}_\text{p}}{\text{d}t} &= \tau_\text{p} \hat{u}_\text{p} + \xi \varphi, \quad \hat x_\text{p}(0) = \hat{x}_\text{p}^0;
    \label{eq: velocity_characteristics_X} \\ 
    \frac{\text{d}\hat{u}_\text{p}}{\text{d}t} &= -\hat{u}_\text{p} + \xi \phi, \quad \hat u_\text{p}(0) = \hat{u}_\text{p}^0;
    \label{eq: velocity_characteristics_U} \\
    \text{and} \nonumber \\
    \begin{split}
    \frac{\text{d}{f}_{X U \Xi}}{\text{d}t} &= {f}_{X U \Xi}, \\
    f_{X U \Xi}(& \hat{x}_\text{p}(0),\hat{u}_\text{p}(0),\xi;0) = \delta(\hat{x}_\text{p}^0) \delta(\hat{u}_\text{p}^0-v_0) f_\Xi(\xi).
    \end{split}
    \label{eq: velocity_characteristics_f}
\end{align}
\end{subequations}
The joint PDF changes along time because the non-conservative version of the equation~\eqref{eq: velocity_Liouville_3D} is non-homogeneous. This occurs because the third term in~\eqref{eq: velocity_Liouville_3D} has an explicit dependency with the particle velocity which leads to a non zero right hand side in the ODE for the joint PDF~\eqref{eq: velocity_characteristics_f}. 
The solution to~\eqref{eq: velocity_characteristics} is 
\begin{subequations}\label{eq: velocity_characteristics_solution}
\begin{align}
    \hat{x}_\text{p}(t) &= \hat{x}_\text{p}^0 + \tau_\text{p} \hat{u}_\text{p}^0 (1 - \text{e}^{-t} ) + \xi \sigma_X, \quad 
    \hat{u}_\text{p}(t) = \hat{u}_\text{p}^0 \text{e}^{-t} + \xi \sigma_U,
    \label{eq: velocity_characteristics_solution_Up} \\
    f_{X U \Xi} & = f_{X U \Xi}^0 \text{e}^{t} .
    \label{eq: velocity_characteristics_solution_f} 
\end{align}
\end{subequations}
where $\sigma_X$ and $\sigma_U$ are defined by~\eqref{eq: velocity_sigmas}. They are derived with the moment model~\eqref{eq: moment_models_q}.
The general expression for the joint PDF $f_{X U \Xi}(x_\text{p},u_\text{p},\xi;t)$ is found  by obtaining $\hat{x}_\text{p}^0$ and $\hat{u}_\text{p}^0$ from the characteristics~\eqref{eq: velocity_characteristics_solution_Up} and substituting them into~\eqref{eq: velocity_characteristics_solution_f}:
\begin{align}
    f_{X U \Xi} = e^{t}f_\Xi(\xi) \delta(  \hat{x}_\text{p}^0) \delta (  \hat{u}_\text{p}^0 - v_0),
\end{align}
which leads to~\eqref{eq: velocity_solution_general_f_axpup}.

The characteristics of~\eqref{eq: FHHS_Liouville_1} are
\begin{align}
    \frac{\text{d}\hat{u}_\text{p}}{\text{d}t} &= -\frac{1}{\tau_\text{p}} \hat{u}_\text{p}+ \xi\phi, \\
    \frac{\text{d}{f}_{U\Xi}}{\text{d}t} &= \frac{1}{\tau_\text{p}} {f}_{U\Xi}, 
\end{align}
that one can be solve analytically, obtaining 
\begin{subequations}\label{eq: FHHS_charac}
\begin{align}        
    \hat{u}_\text{p} &= \hat{u}_\text{p}^0 \text{e}^{- t / \tau_\text{p}} + \xi ( 1 - \text{e}^{-\mathcal{C}_1 t / \tau_\text{p}}  )^{\mathcal{C}_2} , 
    \label{eq: FHHS_charac_Up} \\
    f_{U\Xi} & = f_{U\Xi}^0 \, \text{e}^{t / \tau_\text{p} }, 
    \label{eq: FHHS_charac_f}
\end{align}
\end{subequations}
which leads to~\eqref{eq:sol3}.




\bibliographystyle{unsrtnat}
\bibliography{BIB}  

\end{document}